\documentclass{article}

\usepackage{a4wide}

\usepackage{times}
\usepackage{amsmath}
\usepackage{amssymb}
\usepackage{graphicx}
\usepackage{epsfig}

\newcommand{\I}{\mathbb {I}}

\def\ket#1{| #1 \rangle}

\title{\vspace{-2.5cm} Quantum Metropolis Sampling}

\author
{K. Temme $^{1}$, T.J.\ Osborne$^{2}$,  K. Vollbrecht$^{3}$,  D. Poulin$^{4}$, and F. Verstraete$^{1}$ \\
\\
\normalsize{$^{1}$Fakult\"at f\"ur Physik, Universit\"at Wien, Boltzmanngasse 5, 1090 Wien, Austria}\\
\normalsize{$^{2}$Inst. f. Theoretical Physics, Leibniz Universit\"at Hannover,
Hannover, Germany}\\
\normalsize{$^{3}$Max Planck Institut f\"ur Quantenoptik, Garching, Germany}\\
\normalsize{$^{4}$D\'epartement de Physique, Universit\'e de Sherbrooke, Qu\'ebec, Canada}
}

\date{}

\begin{document}

\maketitle

{\bf The original motivation to build a quantum computer came from Feynman \cite{feynman:1982a}
who envisaged a machine capable of simulating generic quantum mechanical systems,
a task that is believed to be intractable for classical computers. Such a machine would have
a wide range of applications in the simulation of many-body quantum physics, including condensed matter physics,
chemistry, and high energy physics. Part of Feynman's challenge was met by Lloyd \cite{lloyd:1996a}
who showed how to approximately decompose the time-evolution operator of interacting quantum particles
into a short sequence of elementary gates, suitable for operation on a quantum computer.
However, this left open the problem of how to simulate the equilibrium and static properties of quantum systems.
This requires  the preparation of ground and Gibbs states on a quantum computer. For classical systems,
this problem is solved by the ubiquitous Metropolis algorithm \cite{Metropolis:1953a},
a method that basically acquired a monopoly for the simulation of interacting particles.
Here, we demonstrate how to implement a quantum version of the Metropolis algorithm on a quantum computer. This algorithm permits to sample
directly from the eigenstates of the Hamiltonian and thus evades the sign problem present in classical simulations.
A small scale implementation of this algorithm can already be achieved with today's technology.}

\newpage

\section{Introduction}

Since the early days of quantum mechanics, it has been clear there
is a fundamental difficulty in studying many-body quantum systems:
the configuration space -- Hilbert space -- of a collection of particles
grows exponentially with the number of particles. Many of the important
breakthroughs in quantum physics during the 20th century have resulted
from efforts to address this problem, leading to fundamental theoretical
and numerical methods to approximate solutions of the many-body
Schr\"odinger equation. However, most of these methods are limited to weakly interacting
particles; unfortunately, it is precisely when the interactions are strong that
the most interesting physics arises. Notable examples include high-$T_c$ superconductors,
electronic structure in large molecules, and quark confinement in quantum chromodynamics.

The configuration-space explosion problem is not unique to quantum
mechanics: the task of simulating interacting \emph{classical}
particles is challenging for the same reason. It was only with the
advent of  computers in the 1950's that a systematic way of
simulating classical many-body systems was made possible. In their
seminal paper \cite{Metropolis:1953a} Metropolis {\it et al.} devised a
general method to calculate the properties of any substance comprising
individual molecules with classical statistics. This landmark paper is a cornerstone in the simulation
of interacting systems and has had a huge influence on a wide
variety of  fields (see e.g. \cite{durrQCD,geman:1984a,kirkpatrick:1983a}).
The Metropolis method can also be used to simulate certain quantum systems
by a ``quantum-to-classical map'' \cite{suzuki:1987a}. Unfortunately, this quantum
Monte Carlo method is only scalable when the mapping conserves the positivity
of the statistical weights, and fails in the case of fermionic  systems due to the
infamous sign-problem.

As the reality of quantum computers comes closer, it is crucial to
revisit the original motivation of Feynman for building a quantum simulator
and to develop a general method, suitable for quantum computing machines, to calculate the
properties of any substance comprising interacting quantum molecules.
Such an algorithm would have a multitude of applications. In quantum chemistry,
it could be used to compute the electronic binding energy as a function of the coordinates of the nuclei,
thus solving the central problem of interest. In condensed matter physics, it could e.g. be used
to characterize the phase diagram of the Hubbard model as a function of filling factor, interaction strength, and
temperature. Finally, it could conceivably be used to predict the mass of elementary particles,
solving a central problem in high energy physics.

The seminal work of Lloyd \cite{lloyd:1996a} demonstrated that a quantum computer can reproduce
the dynamical evolution of any quantum many-body system. It did not address, however, the crucial
problem of initial conditions: how to efficiently prepare the quantum computer in a state of physical
interest such as a thermal or ground state. Ground states could in principle be prepared using the
quantum phase estimation algorithm \cite{abrams:1999a,aspuru-guzik:2005a}, but this method is in
general not scalable, because it requires a variational state with a large overlap with the ground state.
Methods are known for systems with frustration free interactions \cite{VerstraeteNP} or systems
that are adiabatically connected to trivial Hamiltonians \cite{farhi:2001a}, but such conditions are not
generically satisfied. Terhal and Divincenzo \cite{terhal:2000a} suggested two approaches of how a
quantum computer could sample from the thermal state of a system. The first suggestion is also related to the metropolis
rule, yet left open the problem of how one could get around the no-cloning result and could construct local
updates which can be rejected. This shortcoming immediately leads to an exponential running time of the algorithm, as
already discussed in their paper. The second approach of preparing thermal states is by simulating the system's
interaction with a heat bath. However, this procedure seems to produce rather large errors when run on a
quantum computer with finite resources, and a precise framework to describe these errors seems to be out of reach.
Moreover, certain systems like polymers \cite{Binder}, binary mixtures  \cite{Luijten} and critical spin chains
\cite{Swendsen,Evertz} experience extremely slow relaxation when put into interaction with a heat bath.
The Metropolis dynamics solve this problem by allowing transformations that are not physically achievable,
speeding up relaxation by many orders of magnitude and bridging the microscopic and relaxation time scales;
this freedom is to a large extent responsible for the tremendous empirical success of the Metropolis method.

In this paper we propose a direct quantum generalization of the classical Metropolis algorithm and show
how one iteration of the algorithm can be implemented in polynomial time on a quantum computer.
Our quantum algorithm is not affected by the aforementioned sign problem and can be used to prepare
ground and thermal states of generic quantum many-body systems, bosonic and fermionic.
Like the classical Metropolis algorithm, the quantum Metropolis algorithm is not expected
to reach the ground state of an arbitrary Hamiltonian in polynomial time. The ability to prepare
the ground state of a general Hamiltonian in polynomial time would allow to solve QMA-complete problems.
However, as a rule of thumb it always seems possible to define an update strategy for which the Metropolis algorithm thermalizes efficiently if the physical system thermalizes in polynomial time. There are no obvious reasons why the same should not be true for the quantum Metropolis algorithm.
It also inherits all the flexibility and versatility of the classical method, leading, for instance, to a
quantum generalization of simulated annealing  \cite{kirkpatrick:1983a}.

\section{Summary of results}

In this section, we present a sketch of how the quantum Metropolis algorithm works. Details and generalizations will be worked out in later sections. 

To set the stage for the quantum Metropolis algorithm, let us first
recall the classical version. We can assume for definiteness that the system is composed of $n$ two-level particles, i.e.,
Ising spins.  A lattice of $100$ spins has $2^{100}$ different configurations, so it is inconceivable to average them all.
The key insight of Metropolis \emph{et.\ al.}\ was to  set up a rapidly mixing \emph{Markov chain} obeying detailed balance
that samples from the configurations with the most significant probabilities. This can be achieved by randomly transforming
an initial configuration to a new one (e.g. by flipping a randomly selected  spin): if the energy of the new configuration is
lower than the original, we retain the move, but if the energy is larger, we only retain the move with probability
$\exp\left(\beta (E_{old}-E_{new})\right)$, where $E$ is the energy of the configurations and $\beta$ the inverse temperature.

The challenge we address is to set up a similar process in the quantum case, i.e., to initiate an ergodic  random walk on the
eigenstates of a given quantum Hamiltonian with the appropriate Boltzmann weights. In analogy to a spin flip,  the random walk
can be realized by a random local unitary, and the \emph{move} should be accepted or rejected following the Metropolis rule.
There are, however, three obvious complications: 1) We do not know what the eigenvectors of the Hamiltonian are
(this is precisely one of the problems that we want to solve). 2) Certain operations, such as energy measurements,
 are fundamentally irreversible in quantum mechanics, but the Metropolis method requires rejecting, hence undoing,
 certain transformations. 3) One has to devise a criterion that proves that the fixed point of the quantum random walk is the Gibbs state.

\begin{figure}[b]
\begin{center}
{
\includegraphics*{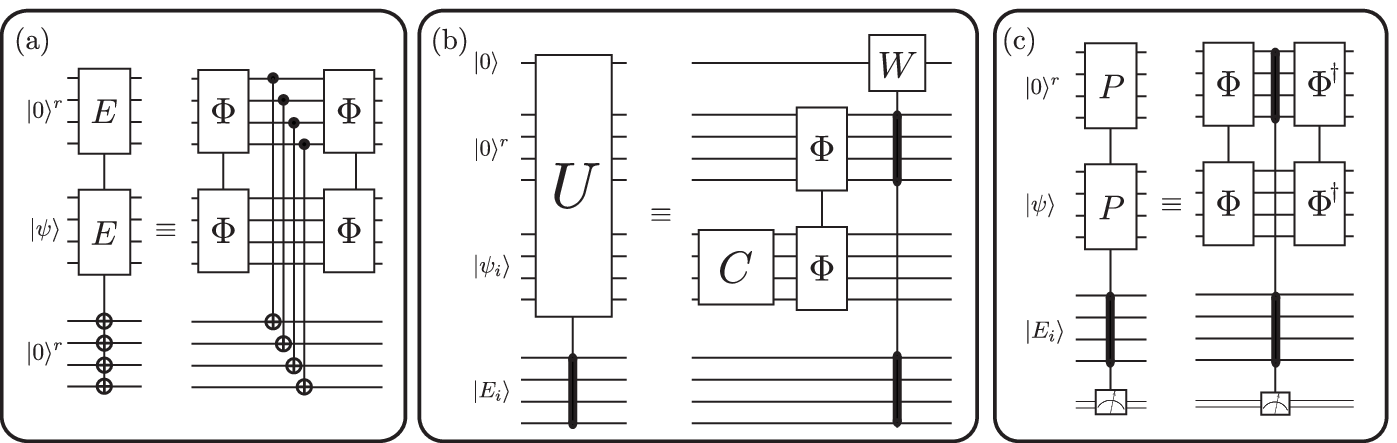}
}
\end{center}
\caption{Fig. (a) The first step of the quantum circuit: the input is an arbitrary state $|\psi\rangle$ and two $r$-qubit registers initialized to $|0\rangle^r$.
Quantum phase estimation $\Phi$ is applied to the state and the second register. The energy value in this register is then copied to the
first register by a sequence of ${\sc cnot}$ gates. An inverse quantum phase estimation is applied to the state and the second register .
Fig. (b) The elementary step in the quantum circuit:  the input is the eigenstate $|\psi_{i}\rangle$ with energy register $|E_i\rangle$ and two registers initialized to $|0\rangle^r$ and $|0\rangle$. The unitary $C$ is then applied, followed by a quantum phase estimation step and the coherent Metropolis gate $W$. The state evolves as follows: $|\psi_{i}\rangle|E_i\rangle|0\rangle|0\rangle\rightarrow C|\psi_{i}\rangle|E_i\rangle|0\rangle|0\rangle=\sum_k x^i_k|\psi_{k}\rangle|E_i\rangle|0\rangle|0\rangle\rightarrow\sum_k x^i_k|\psi_{k}\rangle|E_i\rangle|E_k\rangle|0\rangle\rightarrow \sum_k x_k^i\sqrt{f_k^i}|\psi_{k}\rangle|E_i\rangle|E_k\rangle|1\rangle+\sum_k x_k^i\sqrt{1-f_k^i}|\psi_{k}\rangle|E_i\rangle|E_k\rangle|0\rangle$ with $f_k^i=\min\left(1,exp\left(-\beta(E_i-E_k)\right)\right)$.
Fig. (c) The binary measurement checks whether the energy of the state $|\psi\rangle$ is the same as the energy of the original one $|\psi_i\rangle$. This is done by using an extra register containing phase estimation ancillas, a step that checks whether the energy is equal to $E_i$ or not, and finally an undoing of the phase estimation step that preserves coherence.}
\label{fig.elemUnit}
\end{figure}

To address the first obstacle, we assume for simplicity that the Hamiltonian has non-degenerate commensurate eigenvalues $E_i$,
and denote the corresponding eigenvectors $|\psi_{i}\rangle$. In the supplementary material, it is shown that those conditions are unnecessary.
We can make use of the phase estimation algorithm \cite{kitaev:1995a,cleve:1997a,abrams:1999a,berry:2007a}  to prepare a random energy
eigenstate and measure the energy of a given eigenstate.  Then, each quantum Metropolis step (depicted in Fig. \ref{fig.elemUnit}) takes as
input an energy eigenstate $\ket{\psi_i}$ with known energy $E_i$, and applies a random local unitary transformation $C$, creating the
superposition $C|\psi_i\rangle=\sum_k x^i_k|\psi_k\rangle$.  $C$ could be a bit-flip at a random location like in the classical setting, or
some other simple transformation. The phase estimation algorithm is now used in a coherent way, producing $\sum_k x^i_k|\psi_k\rangle\ket{E_k}$.
At this point, we could measure the second register to read out the energy $E_k$ and accept or reject the move following the Metropolis prescription.
However, such an energy measurement would involve an irreversible collapse of the wave function, which will make it impossible to return to the
original configuration in the case of a reject step.

Classically, we get around this second obstacle by keeping a \emph{copy} of the original configuration in
the computer's memory, so a rejected move can be easily undone. Unfortunately, this solution is ruled
out in the quantum setting by the no-cloning theorem \cite{wootters:1982a}.  The key to the solution is
to engineer a measurement that reveals as little information as possible about the new state, and therefore
only slightly disturbs it.  This can be achieved by a measurement that only reveals one bit of
information---accept or reject the move---rather than a full energy measurement.
The circuit that generates this binary measurement is shown at  Fig. \ref{fig.elemUnit}.
It transforms the initial state $\ket{\psi_i}$ into
\[\underbrace{\sum_k x_k^i\sqrt{f_k^i}|\psi_{k}\rangle\ket{E_i}\ket{E_k}}_{|\psi_i^+\rangle}|1\rangle+
\underbrace{\sum_k x_k^i\sqrt{1-f_k^i}|\psi_{k}\rangle\ket{E_i}\ket{E_k}}_{|\psi_i^-\rangle}|0\rangle\]
where $f_k^i=\min\left(1,\exp\left(-\beta(E_k-E_i)\right)\right)$.  The state can be seen as a coherent superposition
of accepting the update or rejecting it. The amplitudes $x_k^i\sqrt{f_k^i}$ correspond exactly to the
transition probabilities $|x_k^i|^2 f_k^i$ of the classical Metropolis rule. The measurement is completed by measuring
the last qubit in the computational basis. The outcome $\ket 1$ will project the other registers in the state $|\psi_i^+\rangle$. Upon obtaining this outcome, we can measure the second register to learn the new energy $E_k$ and use the resulting energy
eigenstate as input to the next Metropolis step.

\begin{figure}
\begin{center}
{
\includegraphics*{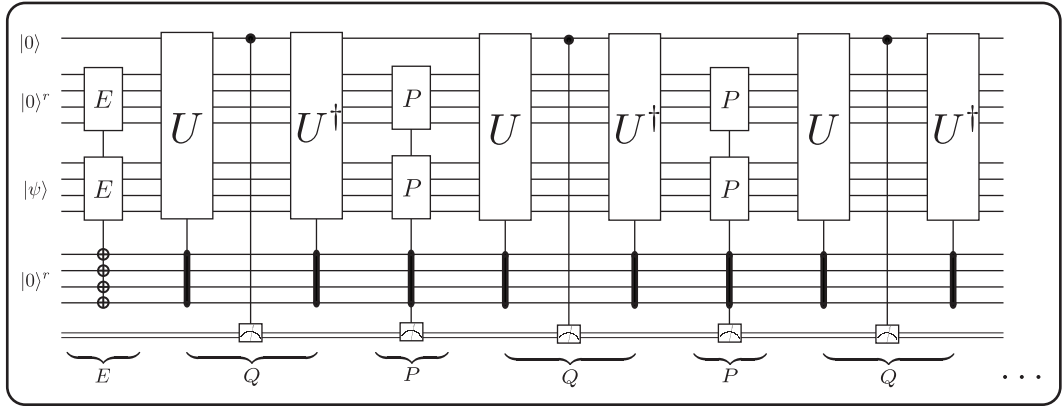}
}
\end{center}
\caption{ The circuit corresponds to a single application of the map ${\cal E}$. 
The first step $E$ prepares an eigenstate of the Hamiltonian, The second step $Q_i$ ,
measures whether we want to accept or reject the proposed update. In the ``reject'' case the 
complete quantum circuit comprises a sequence of measurements of the Hermitian projectors $Q_i$ and $P_i$. 
The recursion is aborted whenever the outcome $P_1$ is obtained, which indicates that we have returned to a 
state with the same energy as the input. Because each iteration has a constant success probability, 
the overall probability of obtaining the outcome $P_1$ approaches 1 exponentially with the number of iterations.}
\label{fig.longCircuit}
\end{figure}

A measurement outcome $\ket 0$ signals that the move must be rejected, so we must return to the input state $\ket{\psi_i}$.
As $|\psi_i^+\rangle$ is orthogonal to to $|\psi_i^-\rangle$ we actually work in a simple 2-dimensional subspace, i.e. a qubit.
In such a case, it is possible to go back to the initial state by an iterative scheme similar to the one employed by Marriott and Watrous
in the context of quantum Merlin Arthur amplification \cite{Watrous}. The circuit implementing this process is shown in Fig.
\ref{fig.longCircuit}. In essence, it repeatedly implements two binary measurements. The first is the one described in the
previous paragraph. The second one, after a basis change, determines if the computer is in the eigenstate $\ket{\psi_i}$
or not.  A positive outcome to the latter measurement implies that we have returned to the input state, completing the rejection;
in the case of a negative outcome, we repeat both measurements. Every sequence of these two measurements has a constant
probability of achieving the rejection, so repeating recursively yields a success probability exponentially close to~1.

The quantum Metropolis algorithm can be used to generate a sequence of $m$ states $|\phi_j\rangle$, $j=1, \ldots, m$
that reproduce the statistical averages of the thermal state $\rho_G = e^{-\beta H}/{\cal Z}$ for any observable $X$:
\begin{equation}
     \frac{1}{m}\sum_{j=1}^m \langle\phi_j|X|\phi_j\rangle = {\rm Tr} X\rho + \mathcal O\left(1/\sqrt{m}\right).
\end{equation}

\begin{figure}
\begin{center}
\scalebox{.75}
{
\includegraphics*{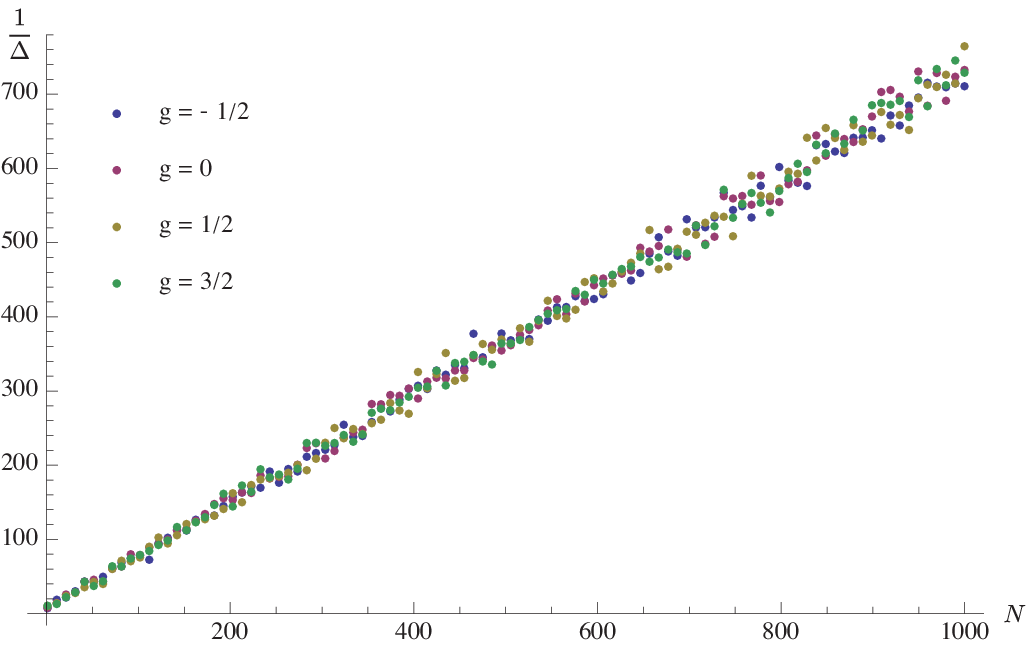}
}
\end{center}
\caption{Inverse gap of the quantum Metropolis map at $T=0$ as a function of the number of spins in a chain with Hamiltonian
$\mathcal{H}=\sum_kX_kX_{k+1}+Y_kY_{k+1}+gZ_k$.  The update rule is a single-spin flip $X_1$; 
remarkably, this single gate is enough to ensure ergodicity. The observed linear scaling indicates that, 
at least in the case of 1D spin chains with nearest - neighbor Hamiltonians,
the quantum Metropolis algorithm converges in polynomial time.}
\label{fig.invGAP}
\end{figure}

To show that the fixed point of the quantum random walk is the Gibbs state, we developed the theory of
{\em quantum detailed balance}.
Let $\{ |\psi_i \rangle \}$ be a  complete basis of the physical Hilbert space  and let  $\{p_i\}$ be a probability distribution on this basis.
Assume that a completely positive map ${\cal E}$ obeys the  condition
\[\sqrt{p_np_m}  \langle{\psi_i} |{\cal E}(|{\psi_n}\rangle \langle {\psi_m}|)|{\psi_j} \rangle
 = \sqrt{p_ip_j}   \langle{\psi_m} |{\cal E}(|{\psi_j}\rangle \langle {\psi_i}|)|{\psi_n} \rangle.\]
 Then $\sigma = \sum_i p_i |{\psi_i} \rangle \langle {\psi_i} |$ is a fixed point of ${\cal E}$.
 The {\em quantum detailed balance} condition only ensures that the thermal state $\rho_G$ is a possible fixed point of the
quantum Metropolis algorithm. The  uniqueness of this fixed point as well as the convergence rate to it
depend on the choice of the set of random unitaries $\{C\}$. If the set of moves are chosen such that the map ${\cal E}$ is ergodic,
the uniqueness of the fixed point is ensured. This condition can be satisfied by choosing $\{C\}$ to be a universal gate set \cite{universal}.
The Metropolis step obeys the {\em quantum detailed balance} condition, if the probability
of applying a specific $C$ is equal to the probability of applying its conjugate $C^\dagger$. This can be seen as the quantum analogue of the
classical symmetry condition for the update probability. In some cases it even suffices to just apply the same local unitary
$C$ at every step of the algorithm (see Fig. ~\ref{fig.invGAP}). In this case, the single unitary $C$ has to be Hermitian and has to ensure ergodicity.
The local unitary can be seen to induce `non-local' transitions between the eigenstates because it is followed
by a phase estimation procedure.

Even though an implementation of this algorithm for full scale quantum many-body problems may be out of
reach for todays technological means, we have presented an algorithm that is indeed scalable to system sizes that are
interesting for actual physical simulations. A small scale implementation of the algorithm
that can be achieved with present day technology is presented in the later sections. Moreover, a discussion is included that sketches the basic
steps necessary for a simulation of some notoriously hard quantum many-body problems. Like in the classical setting the
convergence rate and hence the runtime of the algorithm is dictated by the spectral gap of the stochastic map. The scaling of
the gap depends on the respective problem Hamiltonian and the choice of updates $\{ C \}$.  Just as for the classical Metropolis algorithm,
efficient thermalization is of course not expected for an arbitrary Hamiltonian. This would allow one to solve QMA-complete problems
in polynomial time \cite{Oliveira,Gottesman,SchuchVerstraete}. It is however expected that the algorithm will thermalize if the physical system of interest thermalizes. The inverse gap of the quantum Metropolis map for the XX-chain in a transverse magnetic
field at $T=0$ with a simple single spin flip update as shown in Fig.~\ref{fig.invGAP}.
This plot indicates that the gap scales like $\mathcal{O}(1/N)$ with $N$ the number of spins, even at criticality.
To prove a polynomial scaling of the gap for more complex Hamiltonians remains a challenging open problem.
Also, it is well known that the choice of updates $\{ C \}$  can have a dramatic impact on the convergence rate of the Markov chain
in the classical setting. Finding good updates in the quantum setting is a very interesting open question,
although the above example suggests that the problem might be simpler in the quantum than in the classical case.
The algorithm can be seen as a classical random walk on the eigenstates of the Hamiltonian. All samples are thus computed
with respect to the actual eigenstates. This is why our method is suitable for the simulation of fermionic systems by exploiting the
Jordan - Wigner transformation  \cite{JW} as discussed in \cite{abrams:1997}. The fermionic sign problem is therefore not an
issue for the quantum Metropolis algorithm.  It is worth noting that an additional quadratic speedup might be achievable
using the methods of \cite{szegedy,somma,poulin:2009a}.

\begin{figure}
\begin{center}
{
\includegraphics*{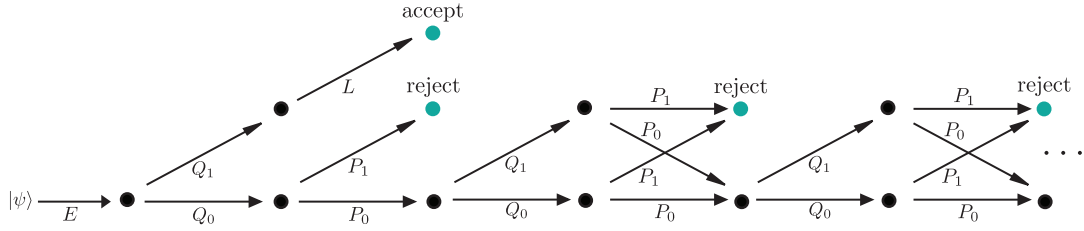}
}
\end{center}
\caption{Given an input state $|\psi\rangle$, we first perform phase estimation to collapse to an eigenstate with known energy $E$.
This graph represents the plan of action conditioned on the different measurement outcomes of the binary $P$ and $Q$ measurements.
Each node in the graph corresponds to an intermediate state in the algorithm.  One iteration of the map is completed when we reach
one of the final leafs labelled by either $\mbox{accept}$ or $\mbox{reject}$. The sequence $E\rightarrow Q_1 \rightarrow L$
corresponds to accepting the update, all other leafs to a rejection.}
\label{Fig.walk}
\end{figure}

\section{Description of the quantum Metropolis algorithm}
\label{mainAlg}

In this section, we provide a more elaborate description of the quantum Metropolis algorithm.
The fundamental building block is the quantum phase estimation algorithm (see section \ref{implementation});
throughout this section we assume that the phase estimation algorithm works perfectly, i.e. given an eigenstate
$|\psi_i\rangle$ of the Hamiltonian $H$ with energy $E_i$, we assume that the quantum phase estimation circuit
$\Phi$ implements the transformation

\[|\psi_i\rangle|0\rangle\rightarrow |\psi_i\rangle|E_i\rangle\]

\noindent where $E_i$ is encoded with $r$ bits of precision. The fact that errors inevitably occur during quantum phase estimation will be dealt with in section \ref{ConvImp}.
The algorithm runs through a number of steps $0..4$ and, just as in the classical case, the total number of iterations of this procedure is related to the autocorrelation times of the underlying stochastic map. As analyzed in the next section, this procedure obeys the quantum detailed balance condition and hence allows to sample from the Gibbs state.  The different steps are also depicted in Fig. \ref{Fig.walk}.

\begin{paragraph}{Step 0:}
Initialize the quantum computer in a convenient state, e.g. $|00\ldots 0\rangle$. We need 4 quantum registers in total. The first one will encode the quantum states of the simulated system, while the other 3 registers are ancillas that will be traced out after every individual Metropolis step. The second register consists of $r$ qubits and  encodes the energy of the incoming quantum state with r bits of precision (bottom register in Fig. \ref{fig.elemUnit}a). The third register is the one used to implement the quantum phase estimation algorithm, also with r qubits  (top register \ref{fig.elemUnit}a). The fourth register is a single qubit that will provide the randomness for accepting or rejecting the Metropolis step.
\end{paragraph}
\begin{paragraph}{Step 1:}
Re-initialize the three ancilla registers and implement the quantum phase estimation based circuit depicted in Fig. \ref{fig.elemUnit}a followed by a measurement of the second register. This prepares an eigenstate $|\psi_{i}\rangle$ with energy $E_i$ and associated energy register $|E_i\rangle$. The upper ancillas are left in the state $|0\rangle^r$ as we assumed perfect phase estimation. The global state is now

\[ |\psi_i\rangle|E_i\rangle|0\rangle|0\rangle\]

\end{paragraph}

\begin{paragraph}{Step 2:}
The next step is depicted in Fig.  \ref{fig.elemUnit}b. Assume that we have defined a set of unitaries ${\cal C} = \{C\}$ that can be implemented efficiently;
those will correspond to the proposed moves or updates  of the algorithm, just like one does for instance spin flips in the case of classical Monte Carlo.
Just as in the classical case, the exact choice of this set of unitaries does not really matter as long as it is rich enough to generate all possible transitions;
the convergence time will, however, depend on the particular choice of moves.  The unitary $C$ is drawn randomly from the set ${\cal C}$ according
to some probability measure $d\mu(C)$. It is only necessary that the probability of choosing a $C$ is equal to the probability of choosing $C^\dagger$,
i.e. $d\mu(C) = d\mu(C^\dagger)$, as this is dictated by the requirement that the process obeys detailed balance, cf. section \ref{perfectPE}.

The new state can be written as a superposition of the eigenstates:

\[C|\psi_{i}\rangle=\sum_k x_k^i|\psi_{k}\rangle\]

\noindent Implement the coherent quantum phase estimation step specified in Fig. \ref{fig.elemUnit}b, which results in the state

\[\sum_k x_k^i|\psi_{k}\rangle\rightarrow  \sum_k x_k^i|\psi_{k}\rangle|E_i\rangle|E_k\rangle|0\rangle.\]

\noindent Note that $E_k$ is only encoded with a precision of $r$ bits, so that in practice there will be a lot of degeneracies.

\noindent Finally, implement the unitary $W(E_k,E_i)$ (Fig. \ref{fig.elemUnit}b) which is a one-qubit operation conditioned on the value of the 2 energy registers:

\begin{eqnarray}\label{Wmatrix}
W(E_k,E_i)&=&\left(\begin{array}{cc} \sqrt{1-f_{ik}} &\sqrt{f_{ik}}\\\sqrt{f_{ik}} & -\sqrt{1-f_{ik}}\end{array}\right) \label{Wki}\\
f_{ik}&=& \min\left( 1, \exp\left(-\beta\left(E_k-E_i\right)\right)\right). \label{Wki2}
\end{eqnarray}

\noindent The system is now in the state

\[\sum_k x_k^i\sqrt{f_k^i}|\psi_{k}\rangle|E_i\rangle |E_k\rangle |1\rangle+\sum_k x_k^i\sqrt{1-f_k^i}|\psi_{k}\rangle |E_i\rangle |E_k\rangle|0\rangle.\]

\noindent For later reference, the product of the three unitaries $C$, the phase estimation step,
and $W$ is called $U$ (see Fig. \ref{fig.elemUnit}b).

\end{paragraph}

\begin{paragraph}{Step 3:}

Measure the single ancilla qubit in the computational basis.  A measurement outcome $1$ corresponds to an
acceptance of the move and collapses the state into

\[\sum_k x_k^i\sqrt{f_k^i}|\psi_{k}\rangle|E_i\rangle|E_k\rangle|1\rangle.\]

\noindent In the case of this accept move, we can next  measure the third register which prepares a new eigenstate
 $|\psi_k\rangle$, and follow that by an inverse quantum phase estimation step. This leads to the state

\[ |\psi_{k}\rangle|E_i\rangle|0\rangle|1\rangle \]

\noindent with probability proportional to  $\left|x_k^i\sqrt{f_k^i}\right|^2$. This state will be the input for the next step
in the iteration of the Metropolis algorithm: go back to  step 1 for this next iteration. Note that the sequence
$E\rightarrow Q_1 \rightarrow L$ depicted in Fig. \ref{Fig.walk} exactly corresponds to this sequence of gates.

A measurement  $|0\rangle$ in the single ancilla qubit signals a reject of the update.
In this case,  first apply the gate $U^\dagger$, and then go to step 4.
\end{paragraph}


\begin{paragraph}{Step 4:}
Let us first define the Hermitian projectors $Q_0$ and $Q_1$, made up of the gates defined in step $2-3$
including the measurement on the ancilla:

\begin{eqnarray*}
Q_0&=&U^\dagger \left(\mathbb{I} \otimes \mathbb{I} \otimes \mathbb{I} \otimes |0\rangle\langle 0|\right)U \\
Q_1&=&U^\dagger \left(\mathbb{I} \otimes \mathbb{I} \otimes \mathbb{I} \otimes |1\rangle\langle 1|\right)U
\end{eqnarray*}

\noindent Let us also define the Hermitian projectors $P_0$ and $P_1$ as

\begin{eqnarray*}
P_0&=&\sum_i\sum_{E_\alpha\neq E_i}|
\psi_{\alpha}\rangle\langle\psi_{\alpha}|\otimes |E_i\rangle\langle E_i| \otimes \mathbb{I} \otimes \mathbb{I}\\
P_1&=&\sum_i\sum_{E_\alpha = E_i}|
\psi_{\alpha}\rangle\langle\psi_{\alpha}|\otimes |E_i\rangle\langle E_i| \otimes  \mathbb{I} \otimes \mathbb{I}
\end{eqnarray*}

\noindent Here equality (or inequality) means that the first $r$ bits of the energies do (not) coincide.
This measurement $P_\alpha$ can easily be implemented by a phase estimation step depicted in Fig. \ref{fig.elemUnit}c.

The fourth step of the algorithm now consists of a sequence of measurements (see Fig. \ref{fig.longCircuit}). First we implement the
von Neumann measurement defined by $P_\alpha$. If the outcome is $P_1$, then we managed to prepare a new eigenstate
$|\psi_{\alpha}\rangle$ with the same energy as the initial one $|\psi_{i}\rangle$, and therefore succeeded in undoing the measurement.
Go to step 1. If the outcome is $P_0$, we do the von Neumann measurement $Q_\alpha$. Independent of the outcome, we again
measure $P_\alpha$, and if the outcome is $P_1$, we achieved our goal, otherwise we continue the recursion (see Fig. \ref{Fig.walk}).
 It happens that  the probability of failure decreases exponentially in the number of iterations (see section \ref{runRej}) ,
 and therefore we have a very good probability of achieving our goal. In the rare occasion where we do not converge
 after a pre-specified number of steps, we abort the whole Monte Carlo simulation and start all over. \\
\end{paragraph}

This finishes the description of the steps in the algorithm.

\subsection{Running time of the rejection procedure:}
\label{runRej}
 Let us discuss the convergence of the reject step more closely.
As already explained, the algorithm should prepare a new state with the same energy as the original one
 $E_i$ in the case of a reject move. As shown in Fig. \ref{Fig.walk}, we will do this by repeating a
 sequence of two different binary measurements $P_i$ and $Q_i$. The  recursion stops, whenever the
 measurement outcome $P_1$ is obtained, where $P_1$ is the projector on the subspace of energy $E_i$.
Note that it is crucial for the algorithm that the initially prepared state  $E | \psi_{i} \rangle|0^{2r+1}\rangle$
is an eigenstate of the projection $P_1$. This is indeed the case, even if we take into account the fluctuations
in the quantum phase estimation step discussed in the next section: the error that is generated by the fluctuations
of the pointer variable can be accounted for if we verify the equality of the energy in $P$ only up to $\tilde{r} < r$
bits of precision. This allows to enlarge the eigenspace of $P$ with approximate energy $E_i$, encompassing
the fluctuations of the pointer variable.\\

Here we will calculate the expected running time. The probability of failure to reject the move,
given that we start in some state $ | \psi_{i} \rangle$ in the energy $E_i$ subspace, after
$n \geq 2$  steps, is given by the probability of measuring $P_0$ after $n$ subsequent binary
measurements. Note that the commutator $[P_0 Q_s P_0, P_0 Q_{s'}P_0] = 0 $ for all $s$, $s'$.
Therefore, see Fig.~\ref{Fig.walk}, the probability of failure can be cast into the form
\begin{eqnarray}
p^{fail}_i(n) = \sum_{m=0}^n \left ( \begin{array}{c} n \\ m \end{array} \right)
\mbox{Tr}\left[ \left(P_0Q_0P_0\right)^{n-m} \left(P_0Q_1P_0\right)^{m} P_0 Q_0 E \right. \\ \nonumber
\left. \left(|\psi_{i}\rangle\langle \psi_{i}| \otimes |0^{2r+1} \rangle \langle 0^{2r+1}| \right)
E Q_0 P_0  \left(P_0Q_1P_0\right)^{m}  \left(P_0Q_0P_0\right)^{n-m} \right].
\end{eqnarray}
The full expression can conveniently be summed up to a single term:
\begin{eqnarray}\label{pfail}
p^{fail}_i(n) = \langle \psi_{i}|\langle 0^{2r+1} |E  Q_0 P_0
 \left[P_0 (\sum_{s=0}^1 Q_s P_0 Q_s) \;P_0\right]^n P_0 Q_0 E |  \psi_{i}\rangle|0^{2r+1}\rangle
\end{eqnarray}
We now make use of the Lemma (\ref{LemJordan}) as stated in section \ref{subSpace} and choose a basis in which
the projectors $P_i$ and $Q_i$ are block diagonal. Note that we reuse the same two pointer registers at
each phase estimation step in the algorithm. This means that even though a realistic phase estimation
procedure does not necessarily act as a projective measurement on the physical subsystem, the
binary measurements $P_i$ and $Q_i$ are still projectors on the full circuit. Therefore Lemma
(\ref{LemJordan}) can still be employed, even for a realistic phase estimation procedure.
Without loss of generality, we assume that the rank of $rank(P_1) = p$ is smaller than the rank of
$Q_1$ which is equal to half the dimension of the complete Hilbert space (note that $P_1$ projects on a
single energy subspace). Assume that the  unitary $U_{J}$ brings $P$ and $Q$ to this desired form.
This allows us to rewrite (\ref{pfail}) as $p^{fail}_i(n) = \langle \psi_i|\langle 0^{2r+1} |E U_{J}^{\dagger}
D_{fail}(n) U_{J} E | \psi_{i} \rangle| 0^{2r+1}\rangle$ with
\begin{eqnarray}
D_{fail}(n) = \left(\begin{array}{cccc} D(\mathbb{I}-D)(D^2 + (\mathbb{I} -D)^2)^n &
-\sqrt{D(\mathbb{I}-D)}(D^2 + (\mathbb{I} -D)^2)^n & 0 & 0  \\ \nonumber
 -\sqrt{D(\mathbb{I}-D)}(D^2 + (\mathbb{I} -D)^2)^n & D^2(D^2 + (\mathbb{I} -D)^2)^n &
 0 & 0 \\ \nonumber 0 & 0 & 1 & 0 \\ \nonumber 0 & 0 & 0 & 1 \end{array} \right).
\end{eqnarray}
Here, $D$ denotes a $p$-dimensional diagonal matrix with only positive entries. Note
that the state\\ $U_{J}E |  \psi_i\rangle| 0^{2r+1}\rangle$ has complete support on the projection operator $P_1$.
That is, as we stated earlier, the state is an eigenstate of $P_1$.  this means that it only acts on the first upper left block.
If we denote by $ 0 \leq d^* \leq 1$ the diagonal entry of $D$ that gives rise to the largest entry in the upper left block of
the matrix $D_\text{fail}(n)$, we can bound
\begin{equation} \label{boundPfail}
p^{\text{fail}}(n) \leq d^*(1-d^*)({d^*}^2+(1-d^*)^2)^n.
\end{equation}
We observe, that the probability of failure decays exponentially in $n$, for a $n$-independent $d^*$.
Let us maximize this expression over all possible values of $d^*$, in order to obtain an
absolute upper bound to the failure probability. Defining $x = {d^*}^2+(1-d^*)^2 = 1-2d^*(1-d^*)$,
we see that this probability may be bounded by $\frac{1-x}{2}x^n$.
This expression is maximized by choosing $ x = \frac{n}{n+1} $, for which we have
\begin{equation}\label{rejBound}
   p_{\text{fail}}(n) \leq \frac{1}{2(n+1)}\left(\frac{1}{1+\frac1n}\right)^n \approx \frac{1}{2e(n+1)}.
\end{equation}
Hence, choosing $n = O(1/\epsilon)$ recursion steps is sufficient to reduce the probability of failure to
below $\epsilon$. We have to choose this $\epsilon$ in such a mannar, that the probability of failure during a
complete cycle of the Metropolis algorithm is bounded by a small constant number.

\subsection{Running time of the quantum Metropolis algorithm}
\label{RunTime}
Let us discuss the runtime scaling of the full Metropolis algorithm. In general, there are
three types of error one has to deal with when we consider the the runtime scaling of the algorithm.\\

First, we are dealing with a Markov chain and hence there is an associated mixing error $\epsilon^{mix}$.
The mixing error of the Markov chain is defined with respect to trace norm distance, as
$\| {\cal E}^{m_{mix}}[\rho_0] - \sigma^*\|_1 \leq \epsilon_{mix}$. Here $m_{mix}$ denotes the mixing time, i.e. the number
of times the completely positive map has to be applied starting from an initial state $\rho_0$ to be $\epsilon_{mix}$
close to the steady state $\sigma^*$ of the Markov chain. The mixing time is determined by the the gap $\Delta$
between the two largest eigenvalues in magnitude of the corresponding completely positive map. The trace norm is bounded by \cite{chi2}
\begin{equation}
	\| {\cal E}^m[\rho] - \sigma^* \|_1 \leq C_{\exp} \left(1 - \Delta \right)^m,
\end{equation}
for a map that obeys quantum detailed balance, where $C_{\exp}$ is some constant that typically scales exponentially in the system size.
 The runtime, or the mixing time, scales therefore as
\begin{equation} \label{mix_scale}
m_{mix} \geq {\cal O}\left(\frac{\ln(1/\epsilon^{mix})}{ \Delta}\right).
\end{equation}
Just as for classical stochastic maps one needs to prove that the gap is bounded by a polynomial in the
system size for each problem instance individually to ensure that the chain is rapidly mixing.
It is generally believed, that to prove rapid mixing for a realistic Hamiltonian is hard.
However, the convergence rate of the classical Metropolis algorithm is in practice favourable
if the physical system thermalizes; this is because the Metropolis steps can mimic the actual
physical thermalization procedure, albeit with the added flexibility of unphysical moves that make
thermalization orders of magnitude faster. It is expected that the same will be true for the quantum Metropolis
algorithm as well. \\
The second type of imperfection relates to the fact, that the reject part of a local move cannot be
implemented deterministically. However, we already showed, cf. \ref{runRej}, that this probability can be made
arbitrary small by increasing the number of iterations in the reject move. For all realistic applications one would
choose a fixed $n^*$ so that one only attempts to perform $n \leq n^*$ reject moves before discarding the sample.
We want to achieve an overall success probability of preparing a valid sample that is bounded by some constant $c$.
What do we mean by that? As already stated the Metropolis algorithm allows one to sample from the  eigenstates
$|\psi_{i} \rangle$ with a given probability $p_i \simeq \exp{(-\beta E_i)}$. Since our reject procedure can only be
implemented probabilistically we have to choose a fixed number of times $n^*$ we try to reject a proposed  update.
The probability of failure $p_{\text{fail}}(n)$ of rejecting a proposed update after $n$ steps is bounded by
$ p_{\text{fail}}(n) \leq \frac{1}{2e(n+1)}$, see (\ref{rejBound}). For the algorithm to work, we want the
algorithm to produce a sample after $m_{mix}$ applications of the map ${\cal E}$ with a probability that is
larger than a constant $c$. Hence the probability of failure after $m_{mix}$ steps should obey
$(1 - p_{\text{fail}}(n^*))^{m_{mix}} \geq c.$ This condition is met if we choose
\begin{equation}
n^* > \frac{m_{mix}}{2e(1-c)}
\end{equation}
This means, that we have to implement for each Metropolis step at most $n^*$ measurements $P_i$ and $Q_i$,
before we discard the sample and start over again. Note that this is a very loose upper bound for the actual number of
reject attempts, since the probability of failure actually decays actually exponentially in $n$, however, with some unknown
constant that is ensured to be smaller than unity. \\
The third error relates to the fact that we are implementing the algorithm on a quantum
computer with finite resources, e.g. a finite register to store the energy eigenvalues in the phase estimation
procedure. This leads to a modification of the completely positive map ${\cal E}$, whose fixed point $\sigma^*$
now deviates from the Gibbs state $\rho_G$ by $\| \sigma^* - \rho_G\|_1 \leq \epsilon^*$.
This error will be discussed in section \ref{ConvImp}.

\section{Fixed point of the algorithm and influence of imperfections}
\label{ConvImp}

In the previous descriptions of the algorithm we only considered the idealized case when we are able to identify
each eigenstate by its energy label. When this is the case, the algorithm can be interpreted as a
classical Metropolis random walk where the configurations of the system are replaced by the eigenstates of a
quantum Hamiltonian. However, this picture falls short if we consider the more realistic scenario of a
Hamiltonian with degenerate energy subspaces. The rejection procedure ensures in this case only
that we end up in the same energy subspace we started from. We therefore need to investigate
the fixed point of the actual completely positive map that is generated by the circuit. We will see that the
quantum Metropolis algorithm yields the exact Gibbs state as its fixed point, if the quantum phase estimation
algorithm resolves the energies of all eigenstates exactly. This is obviously impossible for non integer eigenvalues
as one would need infinitely many bits just to write down the energies in binary arithmetic. However, we will show
that this is not a real problem. A polynomial resolution will yield samples that approximate the Gibbs state very well,
if the Markov chain converges sufficiently fast. For the error analysis we will assume that the ergodicity condition is met,
and that the problem Hamiltonian we are trying simulate is such that the Markov chain is rapidly mixing.
To be precise, for the error analysis we assume that the Markov chain is trace-norm contracting, see section \ref{relPhase}.
We previously discussed the errors that arise due to the finite runtime of the algorithm in section \ref{RunTime} and the error
due to the indeterministic rejection scheme, cf.  section \ref{runRej}. In this section we consider the error that is related
to the implementation of the algorithm. Due to the implementation on a quantum computer three types of error arise.
\begin{enumerate}
\item {\bf Simulation errors.} The quantum phase estimation algorithm requires implementing the dynamics $U = e^{-iHt}$
generated by the system's Hamiltonian for various times $t$. This can only be done within a finite accuracy.
\item {\bf Round-off errors.} The quantum phase estimation algorithm represents the system's energy in binary arithmetic with
$r$ bits. This unavoidably implies that the energy is rounded off to $r$ bits of accuracy.
\item {\bf Phase estimation fluctuations.} As seen in Eq.~\eqref{pointerDist}, given an energy eigenstate of
the system, the quantum phase estimation procedure outputs a random $r$-bit estimate of the corresponding energy.
The output distribution is highly peaked around the true energy, but fluctuations are important and cannot be ignored.
\end{enumerate}
The first error is related to the fact that $\exp(itH)$ has to be approximated by a Trotter-Suzuki unitary.

This error can be ignored as long as the necessary effort in the simulation time $T_H$ to make this small, scales
better than any power of $1/\epsilon_H$ with $\epsilon_H$ being this simulation error \cite{berry:2007a}.
This first source of error can be suppressed at polynomial cost. Another way to tackle this error is to adopt the
analysis done in \cite{poulin:2009a}.

The second type of error is not a problem on its own. Suppose that each eigenvalue of $H$ is replaced by its closest
$r$-bit approximation. The corresponding thermal state would differ from the exact one by factors of $\exp(\beta 2^{-r})$.
By choosing $r \gg \log\beta$, this error can be made arbitrarily small. Note that the simulation cost grows exponentially
with $r$, which implies that our Metropolis algorithm has complexity increasing linearly with $\beta$.

Interestingly, such a problem is  already present in the classical Metropolis algorithm \cite{float},
as one implements the Markov chain on a computer with a floating point error. As a stochastic matrix is non-Hermitean,
a tiny perturbation of the stochastic map (by introducing floating point arithmetic) could in principle change the eigenvectors drastically.
However, nobody ever seems to have encountered such a problem; this might originate from the fact that the detailed
balance condition ensures that the stochastic matrix is well behaved.

The third type of error is more delicate and is intimately related to the second type. Indeed, it is not correct to suppose,
as we did in the previous paragraph, that quantum phase estimation outputs the closest $r$-bit approximation to the energy
of the eigenstate. Rather, it outputs a random energy distributed according to Eq.~\eqref{pointerDist}, sharply peaked around
the exact energy. This distribution can be sharpened by employing a method developed in \cite{nagaj:2009a}: the idea is to
adjoin $\eta+1$ separate pointers, each comprising $r$ qubits, and to perform quantum phase estimation $\eta$ times on the
system using each of the first $\eta$ pointer systems in turn for the readout. Then the \emph{median} of the results in the $\eta$
pointers is computed in a coherent way and written into the $(\eta+1)$th pointer. The probability that the median value deviates from
the true energy by more than $2^{-r}$ is less than $2^{-\eta}$ \cite{nagaj:2009a}. Given an eigenstate of $H$, this leaves two possible
phase estimation outcomes, corresponding to the $r$-bit energy values directly below and directly above the true energy.
Hence, the high confidence phase estimation algorithm acts as
\begin{equation}
|\psi_{i}\rangle|0\rangle \rightarrow |\psi_{i}\rangle \left(\; \alpha_{i}( \lfloor E_i\rfloor ) \; |
\lfloor E_i\rfloor  \rangle + \alpha_{i}( \lceil E_i \rceil )\; |  \lceil E_i \rceil \rangle \;\right) +\mathcal O(e^{-\eta})
\label{pruuuuts}
\end{equation}
where $|\alpha_{i}( \lfloor E_i\rfloor )|^2+|\alpha_{i}( \lceil E_i \rceil )|^2 = 1$ and $\lfloor E_i\rfloor$ and $\lceil E_i \rceil$ are
the two closest $r$-bit approximations to $E_i$. Despite this improvement, it is not possible to make the outcome of  the quantum
phase estimation procedure deterministic. In the worst case where the exact energy for a given eigenstate falls exactly between
two $r$-bit values, the two measurements outcomes will be equally likely. Thus, what we described in the main text as
projectors onto energy bins are not truly von Neumann projective measurements, but rather correspond to generalized
(positive operator valued measure, POVM) measurements on the system.

\begin{paragraph}{Phase estimation unitary and POVM}
To understand this, let us start by writing out the full unitary $\Phi$ of the standard quantum phase estimation
procedure as defined in section \ref{implementation}. The unitary acts on the $N$-qubit register that stores the
state of the simulated system and a single $r$-qubit ancilla register that is used to read out the phase information.
We write
\begin{equation}
\Phi = \sum_{y=0}^{2^r-1} \sum_{x=0}^{2^r-1}  M^{y}_{x} \otimes |x  \rangle \langle y|, \mbox{\;\; where \; \; \;}
M^{y}_{x} = \sum_{j = 1}^{2^N} f(E_j, x-y) |\psi_{j} \rangle \langle \psi_{j}|.
\end{equation}
Note that the function
\begin{equation}\label{otherfunc}
f(E_j , x -y) = \frac{1}{2^r}\frac{e^{i \pi(x - \frac{E_j t}{2\pi} - y)}}{e^{i \frac{\pi}{2^r}(x - \frac{E_j t}{2\pi} - y)}}
 \left(\frac{\sin\left(\pi(x - \frac{E_j t}{2\pi} - y)\right)}{\sin\left(\frac{\pi}{2^r}(x - \frac{E_j t}{2\pi} - y)\right)}\right)
\end{equation}
is complex valued. The operators $M_{x}^{y=0}$ constitute the POVM generated on the system state by the
phase estimation procedure. The label $x$ of the POVM denotes the $r$-bit approximation to the energy
generated by the phase estimation procedure, whereas $y$ corresponds to the initial value of the ancilla register.
The map $\Phi$ is therefore the full unitary of the phase estimation procedure. Due to (\ref{otherfunc}) it becomes clear
that the estimate $x$ of the eigenvalue $E_i$ gets shifted by an amount of $y$, if the ancilla register
is not initialized to $y = 0$.
\end{paragraph}

\subsection{The completely positive map}\label{writeCPmap}

We now investigate the actual completely positive map (cp-map) generated by all unitaries and measurements in more detail.
The full map can be understood as an initialization step denoted by $E$ followed by successive $P$ and $Q$ measurements,
as discussed in section \ref{mainAlg} and illustrated in  Fig. \ref{Fig.walk}.  Note that the projectors $Q_i$ depend on the
random unitary $C$. For each application of the map we draw a random unitary $C$ from the set ${\cal C} = \{C\}$
according to the probability measure $d\mu(C)$. We therefore have to average over the set ${\cal C}$.
The cp-map on the system is obtained by tracing out all ancilla registers. As shown in the previous section \ref{mainAlg},
the error obtained by cutting the number of iterations in the reject case to $n^*$ can be made arbitrarily small;
we can therefore approximate the full map as an infinite sum
\begin{eqnarray}\label{fullmap}
{\cal E}[ \rho] &=& \int_{{ \cal C}} \mbox{Tr}_A \left[ L Q_1 E
\left(\rho \otimes |0^{2r+1} \rangle \langle 0^{2r+1}| \right) E Q_1 L^\dagger \right]  \\ \nonumber
&+& \mbox{Tr}_A \left[ P_1 Q_0  E  \left(\rho \otimes |0^{2r+1} \rangle \langle 0^{2r+1}| \right) E Q_0 P_1 \right]  \\ \nonumber
&+& \sum_{n=1}^\infty  \sum_{s_1 \ldots s_n=0}^1 \mbox{Tr}_A \left[ P_1 Q_{s_n} P_0 \ldots P_0 Q_{s_1} P_0 Q_0 E  \right. \\ \nonumber
&&  \left. \left(\rho \otimes |0^{2r+1} \rangle \langle 0^{2r+1}| \right) E Q_0 P_0 Q_{s_1} P_0 \ldots P_0 Q_{s_n} P_1\right] \; d\mu(C).
\end{eqnarray}
The projective measurements $P_{s}$ and $Q_{s}$ are comprised of several individual operations.
We adopt a new notation: an unmarked sum over the indices written as  small Latin letters, e.g. $k_1,p_1,\ldots$
is taken to run over all $2^r$ integer values of the phase estimation ancilla register. The projectors can be written as
\begin{eqnarray}
&&Q_s  = \sum_{k_1, k_2} \sum_{p_1,p_2} \; C^\dagger {M_{k_2}^{p_1}}^\dagger {M_{k_2}^{p_2}}  C \otimes
|k_1 \rangle \langle k_1| \otimes |p_1 \rangle \langle p_2| \otimes R^s(k_1, k_2),\\\nonumber
&&P_0 = \sum_{k_1 \neq k_2} \sum_{p_1,p_2} \; {M_{k_2}^{p_1}}^\dagger {M_{k_2}^{p_2}}
\otimes |k_1 \rangle \langle k_1| \otimes |p_1 \rangle \langle p_2| \otimes \mathbb{I}, \\\nonumber
&&P_1 = \sum_{k_1 = k_2} \sum_{p_1,p_2} \;{M_{k_2}^{p_1}}^\dagger {M_{k_2}^{p_2}}
\otimes |k_1 \rangle \langle k_1| \otimes |p_1 \rangle \langle p_2| \otimes \mathbb{I}.
\end{eqnarray}
As before, we used the convention that the first register contains the physical state of the system. The second register of $r$-qubits corresponds to the
register that stores the eigenvalue estimates of the first phase estimation, the third register is again used for phase estimation and the
last register sets the single condition bit. The last matrix is defined as
\begin{equation}
R^s(k_1,k_2) = W(k_1,k_2)^\dagger |s \rangle \langle s | W(k_1,k_2),
\end{equation}
\noindent with $W$ defined in (\ref{Wki2}). Furthermore, the first operation in the circuit, that prepares an eigenstate and copies its
energy eigenvalue to the lowest register, is denoted by
\begin{equation}
 E = \sum_{k_1, k_2} \sum_{p_1,p_2} \;{M_{k_2}^{p_1}}^\dagger {M_{k_2}^{p_2}}
 \otimes |k_1 \oplus_{r} k_2 \rangle \langle k_1| \otimes |p_1 \rangle \langle p_2| \otimes \mathbb{I} ,
\end{equation}
where $\oplus_{r}$ denotes an addition modulo $2^r$. For notational purposes we introduced another operation
\begin{equation}
	L =  \sum_{k_1, k_2} \sum_{p_1,p_2} \;{M_{k_2}^{p_1}}^\dagger {M_{k_2}^{p_2}} C
 \otimes |k_1 \rangle \langle k_1| \otimes |p_1 \rangle \langle p_2| \otimes W(k_1,k_2).
\end{equation}
A successful measurement of $Q_1$ at the beginning of the circuit, Fig.~ \ref{fig.longCircuit},  followed by the operation
$L$ corresponds to an acception of the Metropolis update and a further clean-up operation that becomes necessary, when
considering a realistic phase estimation procedure. \\

If we define new super-operators $A[\rho]$ and $B_{n}(\{s_n\})[\rho]$, the cp-map on the physical system can be written as
\begin{equation}\label{CPmap:summands}
{\cal E}[ \rho] = A[\rho] + B_{0}[\rho] + \sum_{n=1}^\infty  \sum_{s_1 \ldots s_n=0}^1 B_{n}(\{s_n\})[\rho].
\end{equation}
Here $A$ denotes the contribution to the cp-map that corresponds to the instance, where the
suggested Metropolis move is accepted. Each of the $B_{n}$ correspond to a rejection
of the update after $n+1$ subsequent $Q$ and $P$ measurements. These superoperators can be expressed as follows:
\begin{equation}
A[\rho] = \sum_{k_1,k_2}  \sum_{d,p_1,q_1}  \int_{{ \cal C}} d\mu(C) \; \min\left(1,e^{-\beta \frac{2\pi}{t}(k_2 - k_1)}\right)\;
{M_{k_2}^{d}}^\dagger {M_{k_2}^{p_1}} C {M_{k_1}^{p_1}}^\dagger {M_{k_1}^{0}} \; \rho \;
{M_{k_1}^{0}}^\dagger {M_{k_1}^{q_1}} C^\dagger {M_{k_2}^{q_1}}^\dagger {M_{k_2}^{d}}.
\end{equation}
Furthermore,
\begin{eqnarray}
B_{0}[\rho] &=& \sum_{k_1} \sum_{l_1,r_1}\sum_{d;p_1,p_2;q_1,q_2}  \int_{{ \cal C}} d\mu(C) \; \langle 0 | R^0(k_1,r_1)R^0(k_1,l_1)| 0 \rangle \\\nonumber
&&{M_{k_1}^{d}}^\dagger {M_{k_1}^{p_2}} C^\dagger {M_{l_1}^{p_2}}^\dagger {M_{l_1}^{p_1}} C {M_{k_1}^{p_1}}^\dagger {M_{k_1}^{0}} \; \rho \;
{M_{k_1}^{0}}^\dagger {M_{k_1}^{q_1}} C^\dagger {M_{r_1}^{q_1}}^\dagger {M_{r_1}^{q_2}} C {M_{k_1}^{q_2}}^\dagger {M_{k_1}^{d}},
\end{eqnarray}
and
\begin{eqnarray}
B_{n}(\{s_n\})[\rho] &=& \sum_{k_1} \sum_{d,\{l_{n+1}\};\{r_{n+1}\}} \; \int_{{ \cal C}} d\mu(C) \; g_{k_1}\left(\{s_n\},\{l_{n+1}\},\{r_{n+1}\}\right) \\\nonumber
&& D_{k_1}^d \left( \{l_{n+1}\}\right) \; \rho \; {D_{k_1}^d}^\dagger \left( \{r_{n+1}\}\right).
\end{eqnarray}
The operators $D$ and the scalar function $g$ in the definition of $B(\{s_n\})^{n}$ are given by
\begin{eqnarray}\label{funkyfunc}
g_{k_1}\left(\{s_n\},\{l_{n+1}\},\{r_{n+1}\}\right) = \langle 0 |R^0(k_1,r_1)R^{s_1}(k_1,r_2) \ldots R^{s_n}(k_1,r_{n+1}) \\\nonumber
R^{s_n}(k_1,l_{n+1}) \ldots R^{s_1}(k_1,l_2) R^0(k_1,l_1)| 0 \rangle
\end{eqnarray}
and
\begin{eqnarray}\label{Dopp}
D_{k_1}^d \left( \{l_{n+1}\}\right) = \sum_{\{a_{n+1}\} \neq k_1} \sum_{\{p_{2n}\}}
{M_{k_1}^{d}}^\dagger {M_{k_1}^{p_{2n}}} C^\dagger {M_{l_{n+1}}^{p_{2n}}}^\dagger
{M_{l_{n+1}}^{p_{2n-1}}} C {M_{a_{n+1}}^{p_{2n-1}}}^\dagger {M_{a_{n+1}}^{p_{2n-2}}} C^{\dagger} \dots \\\nonumber
{M_{a_1}^{p_3}}^\dagger {M_{a_1}^{p_2}} C^\dagger {M_{l_1}^{p_2}}^\dagger {M_{l_1}^{p_1}} C {M_{k_1}^{p_1}}^\dagger {M_{k_1}^{0}}.
\end{eqnarray}
This concludes the description of the completely positive map corresponding to one iteration of the Metropolis algorithm.

\subsection{Fixed point of the ideal chain}
\label{perfectPE}

To be able to make statements about the fixed point of this quantum Markov chain, we introduce (see section \ref{quDetB})
a quantum generalization of the detailed balance concept. As for classical Markov chains, this criterion only ensures that the
state with respect to which the chain is detailed balanced is a fixed point. However, it does not ensure that this fixed point is unique.
The uniqueness follows from the ergodicity of the Markov chain \cite{ergoDM,Michael} and thus depends in our case on the choice of updates $\{C\}$,
which can be chosen depending on the problem Hamiltonian. A sufficient (but not necessary)
condition for ergodicity can easily be obtained by enforcing $\{C\}$ to form a universal gate set, as will be shown below.\\

In section \ref{quDetB}  it is shown that a  quantum Markov chain obeys quantum detailed balance, if there exists a
probability distribution $\{p_i\}$ and a complete set of orthonormal vectors $ \{| \psi_i\rangle\}$ for which
\begin{equation}\label{locDetB}
\sqrt{p_np_m}  \langle{\psi_i} |{\cal E}[|{\psi_n}\rangle \langle {\psi_m}|]|{\psi_j} \rangle
 = \sqrt{p_ip_j}   \langle{\psi_m} |{\cal E}[|{\psi_j}\rangle \langle {\psi_i}|]|{\psi_n} \rangle.
\end{equation}
This condition together with the ergodicity of the updates  $\{C\}$ ensures that the unique fixed point of the quantum
Markov chain is
\begin{equation}
 \sigma = \sum_{i=1}^{2^N} p_i|\psi_i\rangle\langle \psi_i|.
\end{equation}

We therefore would like to verify whether condition (\ref{locDetB}) is satisfied when we choose the $p_i$ equal to the Boltzmann
weights of $H$ and the vectors equal to the eigenvectors  $|\psi_{i}\rangle$.\\

The condition (\ref{locDetB}) is linear in the superoperators. We can therefore conclude that, when each of the
summands $A$ and all the $B$'s in (\ref{CPmap:summands}) individually satisfy this condition,
the total cp-map ${\cal E}$ is detailed balanced.

The idealized case would be met if we could simulate a Hamiltonian $H$ with eigenvalues $E_i$ that are
$r$-bit integer multiples of $\frac{2\pi}{t}$, or if we had an infinitely large ancilla register for the phase estimation.
In this case, the operators $M^{p}_{E}$ would reduce to simple projectors $\Pi_{E+p}$ on the energy subspace labeled by
$E+p$. Hence \[{M^{p}_{E}}^{\dagger} M^{q}_{E} =  \delta_{p,q} \Pi_{E+p}.\] Note that the $\delta_{p,q}$ ensures that
after each $P$ and $Q$ measurement the second ancilla register used for phase estimation is again completely disentangled
and returns to its original value.\\

Furthermore, in the special case when the eigenvalues of the Hamiltonian
are non-degenerate the projectors reduce to $\Pi_{E_i} = | \psi_{i} \rangle\langle \psi_{i}|$. In this case it can be seen
that the dynamics of the algorithm reduce to the standard classical Metropolis algorithm that is described by a classical
stochastic matrix that can be computed as \[S_{ij} = \langle \psi_{j}|{\cal E} \left[ | \psi_{i} \rangle\langle \psi_{i}| \right] |
\psi_{j} \rangle.\] For this special case it is obvious that the detailed balance condition is met. \\

Let us now turn to the more generic case, when the energy eigenvalues are degenerate.
We investigate each of the contributions to the completely positive map (\ref{CPmap:summands}).

\paragraph{The accept instance:}
We first investigate the accept instance described by the operator $A[\rho]$.
\begin{equation}\label{detBopA}
A[\rho] = \sum_{E_1,E_2}  \int_{{ \cal C}} d\mu(C) \; \min\left(1,e^{-\beta(E_2 - E_1)}\right)
\;{\Pi_{E_2}} C \; {\Pi_{E_1}} \; \rho \;{\Pi_{E_1}} C^\dagger {\Pi_{E_2}}.
\end{equation}
The detailed balance criterion (\ref{locDetB})  for $p_i = \frac{1}{Z}e^{-\beta {E}_i}$ and $ | \psi_{i} \rangle$ reads
\begin{equation}
\frac{1}{Z}e^{-\beta(E_i + E_j)/2} \langle \psi_{l} |A[ | \psi_{i} \rangle \langle \psi_{j} | ] | \psi_{m} \rangle
 = \frac{1}{Z}e^{-\beta(E_l + E_m)/2} \langle \psi_{j} |A [ | \psi_{m} \rangle \langle \psi_{l} | ] | \psi_{i} \rangle.
\end{equation}
\\
Note that the chain of operators begins with a projector $\Pi_{E_1}$ and ends with a
projector $\Pi_{E_2}$. The detailed balance condition reads therefore
\begin{eqnarray}
&&\frac{1}{Z}e^{-\beta(E_i + E_j)/2}  \int_{{ \cal C}} d\mu(C) \;  \min\left(1,e^{-\beta(E_l - E_i)}\right) \delta_{E_l,E_m}\delta_{E_i,E_j}
\langle \psi_{l}| C |\psi_{i} \rangle \langle \psi_{j}| C^\dagger | \psi_{m} \rangle \\\nonumber
 &=&  \frac{1}{Z}e^{-\beta(E_l + E_m)/2}  \int_{{ \cal C}} d\mu(C) \; \min\left(1,e^{-\beta(E_j - E_m)}\right) \delta_{E_l,E_m}\delta_{E_i,E_j}
 \langle \psi_{j} | C   |\psi_{m} \rangle \langle \psi_{l}|  C^\dagger  | \psi_{i} \rangle.
\end{eqnarray}
\\Due to the fact that $\frac{1}{Z}e^{-\beta E_l} \min \left(1,e^{-\beta(E_i -  E_l)}\right) = \frac{1}{Z}e^{-\beta E_i} \min \left(1,e^{-\beta(E_l -  E_i)}\right)$,
this reduces to\\
\begin{eqnarray}\label{hermUnit}
 \int_{{ \cal C}} d\mu(C) \; \langle \psi_{l} | C | \psi_{i} \rangle \langle \psi_{j} | C^{\dagger} | \psi_{m} \rangle  =
 \int_{{ \cal C}} d\mu(C) \; \langle \psi_{j} | C | \psi_{m} \rangle \langle \psi_{l} | C^{\dagger} | \psi_{i} \rangle,
\end{eqnarray}
where the energies of the eigenstates have to satisfy  $E_l = E_m$ and $E_i = E_j$. \\

One sees that (\ref{detBopA}) is satisfied when the probability measure obeys
\begin{equation}
d\mu(C) =  d\mu(C^\dagger).
\end{equation}
If we consider an implementation that only makes use of a single unitary $C$ for every update, we have to ensure that this
unitary is Hermitian, i.e. $C = C^{\dagger}$. This symmetry constraint on the measure can be seen as the quantum
analogue of the fact, that we need to choose a symmetric update rule for the classical Metropolis scheme.

\paragraph{The reject instance:}
We now turn to the reject case described by the operators $B_{n}(\{s_n\})[\rho]$ .
The rejecting operators also simplify greatly when we consider the case
of perfect phase estimation. After each phase estimation step the second register disentangles
due to the $\delta_{p_l,p_{l+1}}$, we get
\begin{eqnarray}
B_{n}(\{s_n\})[\rho] = \sum_{E} \sum_{\{l_{n+1}\};\{r_{n+1}\}} g_{E}\left(\{s_n\},\{l_{n+1}\},\{r_{n+1}\}\right) \int_{{ \cal C}} d\mu(C) \;
D_{E}^0 \left( \{l_{n+1}\}\right) \rho  {D_{E}^0}^\dagger \left( \{r_{n+1}\}\right).
\end{eqnarray}
The chain of unitaries and measurement operators in the operator $D$ (\ref{Dopp}) reduces to
\begin{equation}
D_{E}^0 \left( \{l_{n+1}\}\right) = \Pi_E C^\dagger \Pi_{l_{n+1}} C \Pi_E^{\perp} C^{\dagger} \dots
\Pi_E^{\perp} C^\dagger  \Pi_{l_{1}} C \Pi_E,
\end{equation}
where $\Pi_E^{\perp}$ is the projector on to the orthogonal complement of energy subspace $E$.
Note that the first and the last projector in each chain of operators is $\Pi_E$. Hence, all elements
\[\langle \psi_{l} | B_{n}(\{s_n\})[ | \psi_{i} \rangle \langle \psi_{j} | ]  | \psi_{m} \rangle\]
vanish, if all energies are not equal $E_l = E_i = E_j = E_m$. We can therefore disregard the
probabilities $p_i$ on either side of the detailed balance equation  (\ref{locDetB}).
The detailed balance condition thus reads
\begin{equation}\label{simpDet}
\langle \psi_{l} | B_{n}(\{s_n\})[|\psi_{i} \rangle \langle \psi_{j}|]| \psi_{m} \rangle
 =  \langle \psi_{j} | B_{n}(\{s_n\})[ |\psi_{m} \rangle \langle \psi_{l}| ] | \psi_{i} \rangle.
\end{equation}
It is important that the function $g_{E}\left(\{s_n\},\{l_{n+1}\},\{r_{n+1}\}\right)$ (\ref{funkyfunc}) is real.
Due to this fact and furthermore, since all the individual operators $R^s(E,k)$ are Hermitian, we may exchange
the ordering of the indices $\{l_{n+1}\},\{r_{n+1}\}$. That is, we may write
\begin{eqnarray}
&&g_{E}\left(\{s_n\},\{l_{n+1}\},\{r_{n+1}\}\right) = g_{E}\left(\{s_n\},\{l_{n+1}\},\{r_{n+1}\}\right)^* \\\nonumber
&=&\langle 0 |R^0(k_1,l_1)^{\dagger}R^{s_1}(k_1,l_2)^{\dagger} \ldots R^{s_n}(k_1,l_{n+1})^{\dagger}
R^{s_n}(k_1,r_{n+1})^{\dagger} \ldots R^{s_1}(k_1,r_2)^{\dagger} R^0(k_1,r_1)^{\dagger}| 0 \rangle  \\\nonumber
&=&g_{E}\left(\{s_n\},\{r_{n+1}\},\{l_{n+1}\}\right)
\end{eqnarray}
Furthermore, since the individual projectors $\Pi_{l_i}$ and $\Pi_{E}^\perp$ are of course Hermitian, we may write
\begin{eqnarray}\label{calcRej}
&&\langle \psi_{l} | B_{n}(\{s_n\})[\psi_{i} \rangle \langle \psi_{j}]| \psi_{m} \rangle \\\nonumber
&=&\hspace{-0.5cm}\sum_{\{l_{n+1}\};\{r_{n+1}\}} \hspace{-0.5cm}g_{E}\left(\{s_n\},\{l_{n+1}\},\{r_{n+1}\}\right)
\int_{{ \cal C}} d\mu(C) \; \delta_{E_l, E_i , E_j , E_m} \langle \psi_{l} | D_{E_l}^0 \left( \{l_{n+1}\}\right)| \psi_{i} \rangle
\langle \psi_{j} |  {D_{E_l}^0}^\dagger \left( \{r_{n+1}\}\right) | \psi_{m} \rangle \\\nonumber
&=&\hspace{-0.5cm}\sum_{\{l_{n+1}\};\{r_{n+1}\}} \hspace{-0.5cm}g_{E}\left(\{s_n\},\{r_{n+1}\}, \{l_{n+1}\}\right)
\int_{{ \cal C}} d\mu(C) \;  \delta_{E_l, E_i , E_j , E_m} \langle \psi_{j} | {D_{E_l}^0}^\dagger \left( \{r_{n+1}\}\right) | \psi_{m} \rangle
\langle \psi_{l} | D_{E_l}^0 \left( \{l_{n+1}\}\right) | \psi_{i} \rangle  \\\nonumber
&=&\langle \psi_{j} | B_{n}(\{s_n\})[ \psi_{m} \rangle \langle \psi_{l} ] | \psi_{i} \rangle.
\end{eqnarray}
The last equality in (\ref{calcRej}) is precisely due to the fact that we can reorder the indices as previously discussed
and that we are dealing with projectors on the energy subspaces.\\

As already said, a possible set of updates that will ensure ergodicity in general is given by choosing $\{C\}$ equal to a universal gate set. So for instance
the set of all possible single qubit unitaries augmented with the $\mbox{CNOT}$ gate would suffice to ensure ergodicity for an arbitrary
Hamiltonian. To show this, we make use of a result proved in \cite{Michael}, Proposition 3. For completeness, we just repeat the part of the
proof  that is relevant to us.

\begin{paragraph}{Primitive maps}
A completely positive map ${\cal E}$ is called primitive if for all states $\rho$ there exists a natural number $m$ so that,
\begin{equation}\label{primitive}
	{\cal E}^m[\rho] > 0.
\end{equation}
This means that ${\cal E}^m[\rho]$ has to be full rank for some $m$. All primitive maps are strongly irreducible,i.e. ergodic.
That is, if $\cal{E}$ is primitive the map has a unique eigenvalue $\lambda({\cal E})$ with
magnitude $|\lambda({\cal E})| = 1$ and a unique fixed point $\sigma^* > 0$ of full rank.
\end{paragraph}

\begin{paragraph}{Proof:}
By contradiction: Assume that ${\cal E}$ is primitive but not ergodic. This means that one of the following holds:
(a) $\sigma^*$ is not full rank; (b) There is another $\tilde{\sigma}^*$ that corresponds to $\lambda = 1$, i.e. the eigenvalue is degenerate;
or (c) there exists another eigenvalue with $|\lambda'| = 1$. If (a) holds the channel can not  be primitive,  since for all $m$ we have
${\cal E}^m[\sigma^*] = \sigma^*$  which is not full rank.  Now, if (b) we will be able to define an $\epsilon = [\lambda_{max}((\sigma^*)^{-1/2}
\tilde{\sigma}^*(\sigma^*)^{-1/2})]^{-1}$ so that $\sigma^* - \epsilon\tilde{\sigma}^* \geq 0$ is not full rank and we are back in case (a). Furthermore,
if (a) and (b) do not hold but (c), the only other eigenvalues of magnitude $1$ can only be a $p$ -th root of unity for some finite natural number $p$.
This implies, however, that assumtion (b) holds for the $p$-th power ${\cal E}^p$, and thus (a) follows.\\
\end{paragraph}

With this Lemma at hand, it is straight forward to proof the uniqueness of the fixed point. All we need show is that the cp-map
${\cal E}$ is primitive.

\begin{paragraph}{Uniqueness of the Fixed point}
If we choose the set of all possible updates $\{ C \}$ equal to a set of universal gates, then the Metropolis Markov chain is ergodic
for all finite $\beta < \infty$.
\end{paragraph}

\begin{paragraph}{Proof:}
If ${\cal E}$ denotes the map defined in (\ref{CPmap:summands}), according to (\ref{primitive}) all we need to show is
that there is an $m$ such that for every $| \psi \rangle$ and every $\rho$  $\langle \psi | {\cal E}^m[\rho] | \psi \rangle > 0$.
Since $\rho$ can always be written as a convex combination of rank 1 projectors it suffices to choose
$ \rho = | \varphi \rangle \langle \varphi | $.  Furthermore we observe that all $B_{n}$ defined in
(\ref{CPmap:summands}) are positive, i.e.
\begin{equation}\label{schiete}
\langle \psi | B_{n}(\{s_i\})[\tilde{\rho}] |  \psi \rangle \geq 0,
\end{equation}
since this expression can always be written as the trace over the product of positive semi-definite operators for any $\tilde{\rho}$ and $|\psi \rangle$,
see (\ref{fullmap}). We can therefore disregard the contributions from the $B_{n}$ and
focus only on the accept instance $A$ of the map ${\cal E}$, since by virtue of (\ref{schiete}) we have
\begin{equation}\label{schiete2}
\langle \psi | {\cal E}^{m}[ | \varphi \rangle \langle \varphi | ] | \psi \rangle \geq \langle \psi | A^{m}[ | \varphi \rangle \langle \varphi | ] | \psi \rangle.
\end{equation}
We can thus write
\begin{eqnarray}\label{schiete3}
&&\langle \psi | A^{m}[ | \varphi \rangle \langle \varphi | ] | \psi \rangle =  \\ \nonumber
&&\int d\mu(C_1) \ldots d\mu(C_m) \sum_{{E_1 \ldots E_{m+1}}} \prod_{i = 1}^m
\min(1,e^{-\beta(E_{i+1} - E_{i})}) \left |\langle \psi |\Pi_{E_{m+1}}C_m \ldots C_1 \Pi_{E_1}  |\varphi \rangle  \right |^2 \\\nonumber
&&\geq e^{-\beta(E_{max} - E_{min})} \int d\mu(C_1) \ldots d\mu(C_m)   F_{\psi,\phi}(C_1, \ldots C_m).
\end{eqnarray}
Here $E_{max}$ and $E_{min}$ denote the largest and the smallest eigenvalues of the problem Hamiltonian $H$ respectively,
and we defined the integrant $F$ as
\begin{equation}
 F_{\psi,\phi}(C_1, \ldots C_m)  = \sum_{{E_1 \ldots E_{m+1}}}  \left |\langle \psi |\Pi_{E_{m+1}}C_m \ldots C_1 \Pi_{E_1}  |\varphi \rangle  \right |^2.
\end{equation}
Note that the prefactor $e^{-\beta(E_{max} - E_{min})}$ does not vanish for all finite $\beta$. Since the integrant $F$ is non-negative, we
only need to proove that $F$ does not vanish. Since we are drawing the $C_1 \ldots C_m$ from a set of universal gates we can always
find a finite $m$, by virtue of the Solovay -- Kitaev theorem \cite{SolKitaev}, so that there
exists a sequence of gates $C_i$ that ensures that there is a sufficiency large overlap between $| \psi \rangle $
and $C_m \ldots C_1 | \psi \rangle $. That is for a given $\epsilon_m$, there exists a sequence of $m$ gates, so that
\begin{eqnarray}\label{dreck}
	\left |\langle \psi | C_m \ldots C_1 |\varphi \rangle  \right |^2  =
	\left |  \sum_{E_1 \ldots E_{m+1}} \langle \psi |\Pi_{E_{m+1}}C_m \ldots C_1 \Pi_{E_1}  |\varphi \rangle \right |^2 \geq 1 - \epsilon_m,
\end{eqnarray}
where we inserted resolutions of the identity $\sum_{E_i} \Pi_{E_i}$. Hence, at least one of summands in  (\ref{dreck})
has to be non-zero and thus $F_{\psi,\varphi}$ is  strictly positive and does not vanish.
Therefore, there exists an integer $m$ so that the integral in the last line of (\ref{schiete3}) is strictly positive.  Since  (\ref{schiete3})
acts as a lower bound to  $\langle \psi | {\cal E}^{m}[ | \varphi \rangle \langle \varphi | ] | \psi \rangle$ we can conclude that ${\cal E}$ is
primitive.
\end{paragraph}

\subsection{Error bounds and realistic phase estimation}
\label{relPhase}
Let us next return to a more general Hamiltonian that has a realistic spectrum. As was discussed earlier, a realistic phase estimation
procedure introduces errors not only due to the rounding of the energy values, but more importantly due to the fluctuations of the
pointer variable. For a completely positive map with realistic  phase estimation the detailed balance condition (\ref{locDetB})
will not be met exactly, but we can show that the condition is satisfied approximately. This will be sufficient for our purposes.\\

In order to bound this error we adopt a standard procedure also used for classical Markov chains \cite{ContractionRef}.
Throughout this analysis we assume that the completely positive map is well behaved and is contracting.
Whether this assumption is satisfied depends on the mixing properties of the problem we consider and on the choice of updates.
Therefore, these properties have to be verified for every problem instance individually.
A quantum Markov chain is trace - norm contracting if it satisfies
\begin{equation}
	\| {\cal E}[\rho - \sigma]\|_{1} \leq \eta_1 \| \rho - \sigma \|_{1},
\end{equation}
where the constant $\eta_1 < 1$ is the smallest constant, so that this inequality holds \cite{ContractionRef}.
The constant $\eta_1$ is often referred to as the ergodicity coefficient. Note that the map is considered contracting
only when the constant is strictly smaller than unity. It can occur, for some pathologically behaved maps, that this constant is not
strictly smaller than unity even though the map is rapidly mixing. However, this can be cured by blocking several applications
of the channel together, leading to a new constant smaller than unity \cite{ErrorRef}.
\begin{paragraph}{Error bound}
The error $\epsilon^*$ between the exact fixed point $\sigma^{*}$ of the map ${\cal E}$ and the
Gibbs state $\rho_G = \frac{1}{Z} \exp{\left(-\beta H\right)}$ can be bounded by
\begin{equation}
\| \sigma^* - \rho_G \| \leq \frac{\epsilon^{sg}}{1 - \eta_1}.
\end{equation}
Here $\eta_1 < 1$ is the ergodicity coefficient of ${\cal E}$ and $\epsilon^{sg}$ the error that arises
due to a single application of the map on $\rho_G$, i.e. $ \| {\cal E}[\rho_G] - \rho_G \|_1 \leq \epsilon^{sg} $ .
\end{paragraph}
\begin{paragraph}{Proof:}
The error $\epsilon^*$ can be written as
\begin{eqnarray}
\| \sigma^* - \rho_G \| &=& \lim_{m \rightarrow \infty}  \| {\cal E}^m[\rho_G] - \rho_G \|_1 \leq
\lim_{m \rightarrow \infty} \sum_{k=1}^{m} \|  {\cal E}^k[\rho_G] - {\cal E}^{k-1}[\rho_G] \|_1 \\\nonumber
&\leq&  \lim_{m \rightarrow \infty} \sum_{k=1}^{m}  \eta_1^{k-1} \|  {\cal E}[\rho_G] -
\rho_G \|_1 = \frac{ \|  {\cal E}[\rho_G] -  \rho_G \|_1}{1 - \eta_1}.
\end{eqnarray}
\end{paragraph}
Thus we only need to bound the error that occurs when we apply the map ${\cal E}$ to the Gibbs state $\rho_G$ once.
In order to bound this error, we will make use of the fact that the completely positive map satisfies the
detailed balance condition (\ref{locDetB}) at least approximately. Let us discuss what it means to satisfy detailed balance approximately.
\begin{paragraph}{Approximate detailed balance}
Suppose we are given a completely positive map ${\cal E}$ and an orthonormal basis $\{ | \psi_i \rangle \}$.
To each state we assign a Boltzmann weight of the form $\{p_i  = \frac{1}{Z} e^{-\beta E_i}\}$.
 If this cp-map does not precisely satisfy detailed balance, but only an approximate form such as
\begin{equation}\label{locDetBbis}
\sqrt{p_np_m}  \langle{\psi_i} |{\cal E}[|{\psi_n}\rangle \langle {\psi_m}|]|{\psi_j} \rangle
 = \sqrt{p_ip_j}   \langle{\psi_m} |{\cal E}[|{\psi_j}\rangle \langle {\psi_i}|]|{\psi_n} \rangle \left(1+\mathcal{O}(\epsilon^{sg})\right),
\end{equation}
we can give the following bound on the error, measured in the trace - norm, that occurs upon a single application of the
completely positive map.
\begin{equation}
\|{\cal E}[\rho_G] - \rho_G\|_{1} \leq {\mathcal O}(\epsilon^{sg})
\end{equation}
\end{paragraph}

\begin{paragraph}{Proof:}
Let us define $\rho = \sum_i p_i |\psi_i \rangle \langle \psi_i |$. Then due to (\ref{locDetBbis}) we have
\begin{eqnarray}
\langle \psi_l | {\cal E}[\rho_G] | \psi_m \rangle =  \sum_i p_i  \langle \psi_l | {\cal E}[|\psi_i \rangle \langle \psi_i |] | \psi_m \rangle = \\\nonumber
\sqrt{p_l p_m}  \left(1+\mathcal{O}(\epsilon^{sg})\right) \mbox{Tr}\left [{\cal E}[|\psi_m \rangle \langle \psi_l |] \right] =
p_m \left(1+\mathcal{O}(\epsilon^{sg})\right) \delta_{ml}.
\end{eqnarray}
So the application of ${\cal E}$ yields ${\cal E}[\rho_G] = \tilde{\rho_G}$. Note that the state $\tilde{\rho_G}$ is still diagonal in the same
basis as $\rho_G$ and  both of the probabilities $\tilde{p}_i$ of  $\tilde{\rho_G}$ relate to the original probabilities via
$ \tilde{p}_i = p_i \left(1+\mathcal{O}(\epsilon^{sg})\right)$. Since  $\rho_G$ and $\tilde{\rho_G}$ are both diagonal in the same basis,
it is straightforward to compute that $ \|\tilde{\rho_G} - \rho_G\|_{1} \leq {\mathcal O}(\epsilon^{sg})$.
\end{paragraph}
\\\\

Let us now verify the approximate detailed balance condition (\ref{locDetBbis}) of the completely positive map (\ref{CPmap:summands})
for a realistic spectrum of the Hamiltonian $H$. First let us consider the standard phase estimation procedure.
Since the actual eigenvalues may have arbitrary real values, we may not assume that the individual
$M^y_x$ act as projectors on the system. Note that even the combination of ${M_k^p}^\dagger M_k^q$ is not Hermitian
anymore when $p \neq q$. This is precisely due to the fact that the function $f(E_j,k-p)$  (\ref{otherfunc}) is
complex valued. An additional phase is imprinted on the system state. At first sight this seems
to hinder any form of detailed balance in the eigenbasis of the Hamiltonian.  It turns out, however,
that the total expression on either side of the detailed balance equation is still real.
Note that ${M_k^p}^\dagger M_k^q$ is diagonal in the eigenbasis of $H$ and  assumes the form
\begin{equation}
 {M_k^p}^\dagger M_k^q = \sum_{j=1}^{2^N} f(E_j,k-p)^* f(E_j,k-q) |\psi_j \rangle \langle \psi_j | .
\end{equation}
Hence, the phases  in $f(E_j,k-p)^* f(E_j,k-q)$ cancel up to a total phase factor
$\frac{e^{i \pi(p - q)}}{e^{i \frac{\pi}{2^r}(p-q)}}$, which is independent of both $k$ and $E_j$.
This allows us to write
\begin{eqnarray}
{M_k^p}^\dagger M_k^q  \equiv \frac{e^{i \pi(p - q)}}{e^{i \frac{\pi}{2^r}(p-q)}} S^{pq}_k,
\end{eqnarray}
where now ${S^{pq}_k}^\dagger = {S^{pq}_k}$. Let us have look at a segment of the chain of operators as they
typically appear in the superoperators $A$ or $B$ (\ref{CPmap:summands}). The typical sequences look like
\begin{equation} \label{phasescancel}
\ldots {M_{k_2}^{p_3}}^\dagger{M_{{k_2}}^{p_{2}}}\; C \; {M_{k_1}^{p_{2}}}^\dagger M_{k_1}^{p_{1}}
\ldots \;\; \rightarrow \;\; \ldots \frac{e^{i \pi(p_3 - p_1)}}{e^{i \frac{\pi}{2^r}(p_3-p_1)}} \;
S_{k_2}^{p_3p_2} \;C \; S_{k_1}^{p_2p_1}\ldots
\end{equation}
This leads us to the conclusion that in each of the operator sequences the phases that arise due do to imperfect phase procedure cancel.
The first phase associated to $p_0$ is $0$ due to the initialization, whereas the last phase associated with $d$ is canceled due to the measurement.
This gives an additional explanation of why it is necessary to reuse the same pointer register for the phase estimation procedure each time.
However, this comes at a cost as the realistic phase estimation procedure doesn't naturally disentangle the pointer
register used for the next phase estimation anymore. Hence, the initial state of the ancilla register for the next phase estimation step may be altered.
So after subsequent measurements using the same register the distribution function of the pointer variable spreads.\\
\\
We now consider what happens in the case where we use the high confidence phase estimation based on the \emph{median} - method \cite{nagaj:2009a}.
As already stated, this method allows us to perform phase estimation where the pointer variable fluctuates at most in the order of $2^{-r}$.
All other fluctuations are suppressed by a factor of $2^{-\eta}$ and will therefore be neglected in the following.
According to (\ref{pruuuuts}) we can replace the function $f(E_j,k-p)$ by its enhanced counterpart
$\alpha_{E_j}(k-p)$, which acts as a binary amplitude for the two closest $r$-bit integers to the actual energy $E_j$. As discussed earlier,
the phases that arise due to the imperfect phase estimation algorithm cancel, if for each of the $\eta$ phase estimations the corresponding registers are reused.
We are therefore left again with operators $S^{pq}_k$ acting on the physical system that are diagonal and have only real entries. We will thus regard the amplitudes
 $\alpha_{E_i}(k-p)$ as real from now on. We will therefore write
 \begin{equation}
 	S^{pq}_k = \sum_{j=1}^{2^N} \alpha_{E_j}(k-p) \alpha_{E_j}(k-q) |\psi_j \rangle \langle \psi_j |.
 \end{equation}
Let us pause for a minute and have a closer look at the operators $S^{pq}_k$.
As stated previously  the  $S^{pq}_k$ are diagonal in the Hamiltonians eigenbasis and
have only real entries. Hence, these operators are Hermitian. Furthermore, since $\alpha_{E_j}^2$ acts as a binary probability distribution on the two $\delta = 2^{-r}$ closest
integers to $\frac{E_j t}{2\pi}$, we see that for a fixed $E_j$ and a fixed $q$,  the only possible two values for $k$ are
\[k^{\uparrow} = \left \lceil \frac{E_j t}{2\pi} \right \rceil_{2^{-r}} + q  \mbox{\;\;and\;\;}  k^{\downarrow} = \left \lfloor \frac{E_j t}{2\pi} \right \rfloor_{2^{-r}} + q.\]
Conversely, the operator $S^{pq}_k$ has only support on the subspace spanned by the eigenvectors $|\psi_j \rangle$ whose energies lie in the interval
\[ E_j \in \left [(k+q) - 2^{-r} ; (k+q) + 2^{-r} \right] \cap \left[ (k+p) - 2^{-r} ; (k+p) + 2^{-r} \right].\]
This allows a further conclusion. For a fixed $k$ and $q$ the operator does not vanish only if
\[ p \in [q -2^{-r+1};  q + 2^{-r+1}].\]
The interpretation is as follows: the operator $S^{pq}_k$ implements the action of a phase estimation and its conjugate on the system.
If the ancilla register was initially in the state $|q\rangle$ the full phase estimation process does not disentangle the ancilla register
afterwords, if we have performed in an intermediate operation. We have seen previously in the analysis for the idealized phase estimation
procedure, see section \ref{perfectPE}, that the inverse phase estimation procedure returns the ancilla register to its
original value $|q\rangle$. Since the pointer variable fluctuates now, this is not the case anymore and the pointer register remains
entangled with the simulated system. However, since we perform an enhanced phase estimation procedure, the allowed values for
the ancilla register are bounded by $p^{\pm} = q \pm 2^{-r+1}$. Thus even though $S^{pq}_k$ is not a projector anymore,
the previously discussed conditions suffice to ensure approximate detailed balance.
\\

Let us now verify the approximate detailed balance condition for each of the summands in (\ref{CPmap:summands}).
\paragraph{The accept instance:}
We analyze what happens in the accept case indicated by the operator $A[\rho]$.
Due to the cancellation of the spurious phases (\ref{phasescancel}) this operator
has the form
\begin{equation}
A[\rho] = \sum_{k_1,k_2}  \sum_{d,p_1,q_1} \int_{{ \cal C}} d\mu(C) \; \min\left(1,e^{-\beta \frac{2\pi}{t}(k_2 - k_1)}\right)\;
{S_{k_2}^{d p_1}} C \; {S_{k_1}^{p_1 0}} \; \rho \;{S_{k_1}^{ 0 q_1}} C^\dagger {S_{k_2}^{q_1 d}}.
\end{equation}
We now want to verify whether the approximate detailed balance condition is met, when we
choose again $p_i = \frac{1}{Z}^{-\beta E_i}$ and $|\psi_i \rangle$ as the eigenstate of $H$.
We choose a symmetric measure, i.e. $d\mu(C^\dagger) = d\mu(C)$, and verify the approximate
detailed balance condition (\ref{locDetBbis}). The left side of the equation reads
\begin{eqnarray}
&&\frac{1}{Z}e^{-\beta(E_i + E_j)/2} \langle \psi_{l} |A[|\psi_{i} \rangle \langle \psi_{j}|]| \psi_{m} \rangle \\\nonumber
&=&\sum_{k_1,k_2}\sum_{d,p_1,q_1}  \frac{1}{Z}e^{-\beta(E_i + E_j)/2} \int_{{ \cal C}} d\mu(C) \; \min\left(1,e^{-\beta \frac{2\pi}{t}(k_2 - k_1)}\right)\;
\langle\psi_{l} | {S_{k_2}^{d p_1}} C \; {S_{k_1}^{p_1 0}}  |\psi_{i} \rangle
\langle \psi_{j}|{S_{k_1}^{ 0 q_1}} C {S_{k_2}^{q_1 d}}| \psi_{m} \rangle \\\nonumber
&=& \sum_{k_1,k_2}\sum_{d,p_1,q_1}  \frac{1}{Z}e^{-\beta(E_i + E_j)/2}  \int_{{ \cal C}} d\mu(C) \; \min\left(1,e^{-\beta \frac{2\pi}{t}(k_2 - k_1)}\right)\;
\langle\psi_{l} |  C |\psi_{i} \rangle \langle \psi_{j}| C | \psi_{m} \rangle\\\nonumber
&&\alpha_{E_l}(k_2 - d)\alpha_{E_l}(k_2 - p_1) \alpha_{E_i}(k_1 - p_1)\alpha_{E_i}(k_1)
\alpha_{E_m}(k_2 - d)\alpha_{E_m}(k_2 - q_1) \alpha_{E_j}(k_1 - q_1)\alpha_{E_j}(k_1).\\\nonumber
\end{eqnarray}
We are free to relabel all the summation indices $k_1,k_2,d,\ldots$ to match it with the other side of the equation. The sequence
\begin{eqnarray}
k_2 =  k'_1 + d \rightarrow \left \{ \begin{array}{c}  p_1 = q_1' + d  \\  q_1 = p'_1 + d  \end{array} \right\}
\rightarrow k_1 = k'_2 + d \rightarrow d = 2^r - d'
\end{eqnarray}
does exactly this. Note that since $\alpha_{E_j}(k + 2^r) = \alpha_{E_j}(k)$ the constant $2^r$ in the last step can be dropped.
If we now consider the worst case scenario of the fluctuations of $\alpha_{E_i}(k_1)$, we see that $k_1$ deviates at most as much as
$k_1 \approx \frac{E_i t}{2 \pi} \pm 2^{-r+1}$. The same is also true for $k_2$ and $k'_2$,$k'_1$ respectively. Hence we can conclude
\begin{equation}
\frac{1}{Z}e^{-\beta E_i}  \min\left(1,e^{(-\beta \frac{2\pi}{t}(k_2 - k_1))}\right) =
 \frac{1}{Z}e^{-\beta E_l}  \min\left(1,e^{(-\beta \frac{2\pi}{t}(k'_1 - k'_2))}\right)\left(1 + {\cal O}(\beta\frac{4\pi}{t}2^{-r})\right).
\end{equation}
We can therefore establish, that
 \begin{equation} \label{acceptFluc}
 \frac{1}{Z}e^{-\beta(E_i + E_j)/2} \langle \psi_{l} |A[|\psi_{i} \rangle \langle \psi_{j}|]| \psi_{m} \rangle
 = \frac{1}{Z}e^{-\beta(E_l + E_m)/2} \langle \psi_{j} |A[ |\psi_{m} \rangle \langle \psi_{l}| ] | \psi_{i} \rangle \left(1+\mathcal{O}(\epsilon)\right)
 \end{equation}
\noindent with $\epsilon=\beta\frac{4\pi}{t}2^{-r}$ which can be fully controlled by adjusting the relevant free parameters.

 \subparagraph{The reject instance}
 We now turn to the reject case. The operators change accordingly. We consider the detailed balance
 condition for each of the full $B_{n}(\{s_n\})[\rho]$. Note that due to the previously discussed phase cancellations
 the operators $ D_{k_1}^d \left( \{l_{n+1}\}\right)$ as defined in (\ref{Dopp}) assume the form
 \begin{equation}\label{theNewD}
 D_{k_1}^d \left( \{l_{n+1}\}\right) = \sum_{\{a_{n+1}\} \neq k_1} \sum_{\{p_{2n}\}}
{S_{k_1}^{d p_{2n}}} C^\dagger {S_{l_{n+1}}^{ p_{2n} p_{2n-1}}} C {S_{a_{n+1}}^{p_{2n-1} p_{2n-2}}} C^{\dagger} \dots
{S_{a_1}^{p_3 p_2}} C^\dagger {S_{l_1}^{p_2 p_1}} C {S_{k_1}^{p_1 0}}.
 \end{equation}
The analysis of the reject case is very similar to the exact case. We make use of the fact that all the functions
$g_{k_1}\left(\{s_n\},\{l_{n+1}\},\{r_{n+1}\}\right)$ and $\alpha_{E_i}(k-p)$ are real, and that we can relabel
the indices like we did in the exact analysis. We have to establish that
\begin{eqnarray} \label{rejectFluc}
 && \frac{1}{Z}e^{-\beta(E_i + E_j)/2} \langle \psi_{l} |B_{n}(\{s_n\})[|\psi_{i} \rangle \langle \psi_{j}|]| \psi_{m} \rangle \\\nonumber
&=& \frac{1}{Z}e^{-\beta(E_l + E_m)/2} \langle \psi_{j} |B_{n}(\{s_n\})[ |\psi_{m} \rangle \langle \psi_{l}| ] | \psi_{i} \rangle
 \left(1+\mathcal{O}(\epsilon)\right),
 \end{eqnarray}
up to some $\epsilon$, that will turn out to be $\epsilon = n \frac{4 \pi}{t} \beta 2^{-r}$. We again start by considering the left side of
(\ref{rejectFluc}) and show that it will be equal to the right side up the specified $\epsilon$.
\begin{eqnarray}
&& \frac{1}{Z}e^{-\beta(E_i + E_j)/2} \langle \psi_{l} |B_{n}(\{s_n\})[|\psi_{i} \rangle \langle \psi_{j}|]| \psi_{m} \rangle \\\nonumber
&=& \sum_{k_1} \sum_{d;\{l_{n+1}\};\{r_{n+1}\}} g_{k_1}\left(\{s_n\},\{l_{n+1}\},\{r_{n+1}\}\right) \int_{{ \cal C}} d\mu(C) \; \frac{1}{Z}e^{-\beta(E_i + E_j)/2} \\\nonumber
&&\langle \psi_{j} |  {D_{k_1}^d}^\dagger \left( \{r_{n+1}\}\right) | \psi_{m} \rangle \langle \psi_{l} | D_{k_1}^d \left( \{l_{n+1}\}\right) | \psi_{i} \rangle.
\end{eqnarray}
We will first exchange the index sets $\{r_{n+1}\}$ and $\{l_{n+1}\}$. This is possible since the function $g_{k_1}$  is
real and we follow the same analysis we already performed in the case of the idealized phase estimation.
Now we turn to the sequence of the relabeling of the index set $d, k_1, l_1, r_1, a_1 , b_1, \ldots$. Note that $a_i$ and $b_i$
are part of the definition of $D_{k_1}^d \left( \{l_{n+1}\}\right)$ and ${D_{k_1}^d}^\dagger \left( \{r_{n+1}\}\right)$
respectively (\ref{theNewD}). The relabeling sequence that does what we want reads
\begin{eqnarray}
&&k_1 = k'_1 + d  \rightarrow \\\nonumber
&&\left \{ \begin{array}{c}  p_{2n} = q_{2n}' + d    \\  q_{2n} = p'_{2n} + d  \end{array} \right\}   \rightarrow
\left \{ \begin{array}{c}  l_{n+1} = l'_{n+1} + d  \\  r_{n+1} = r'_{n+1} + d  \end{array} \right\} \rightarrow
\left \{ \begin{array}{c}  p_{2n-1} = q_{2n-1}' + d    \\  q_{2n-1} = p'_{2n-1} + d  \end{array} \right\}   \rightarrow \\\nonumber
&&\left \{ \begin{array}{c}  a_{n+1} = b'_{n+1} + d  \\  b_{n+1} = a'_{n+1} + d  \end{array} \right\} \rightarrow \;\; \ldots \;\;
\left \{ \begin{array}{c}  l_{1} = l'_{1} + d  \\  r_{1} = r'_{1} + d  \end{array} \right\} \rightarrow
\left \{ \begin{array}{c}  p_{1} = q_{1}' + d    \\  q_{1} = p'_{1} + d  \end{array} \right\} \rightarrow d = 2^r - d'.
\end{eqnarray}
For these replacements to work, it is important to note that the operators $R^{s}(k_1,l_i)$ depend only on the
differences, i.e. $R^{s}(k_1-l_i)$. The sequence of replacements therefore leaves the function
$g_{k_1}\left(\{s_n\},\{l_{n+1}\},\{r_{n+1}\}\right)$ unchanged. However, since we do perform $2n$
phase estimation processes for each of the superoperators $B_{n}(\{s_n\})$,
the variable $k_1$ in the last process may fluctuate in the order of $n2^{-r+1}$, as was discussed earlier,
and we may no longer assume that the statistical weights on either side of the equation are equal.
Hence we know that for the worst instance $k_1$ is $\delta = \pm n2^{-r+1}$ close to either energy $E_i$ , $E_j$ , $E_l$ , $E_m$.
We can therefore  see, upon evaluating (\ref{rejectFluc}), that the detailed balance condition for each individual $B_{n}$ is met up to an
$\epsilon = n \frac{4 \pi}{t} \beta 2^{-r}$.
\\

We observe that the $\epsilon$ increases linearly in the number $n$ of subsequent $P$ and $Q$ measurements we make to reject the proposed update.
For all realistic applications, as discussed in section \ref{RunTime}, one would choose a fixed $n^*$ so that one only would attempt to perform $n \leq n^*$
reject moves before discarding the sample. Since we want to achieve an overall success probability of preparing a valid sample that is lower
bounded by a constant $c$, we have to choose $n^* > \frac{m}{2e(1-c)}$. Here $m$ denotes the number of times we have to apply the map
${\cal E}$ to be sufficiently close to the desired steady-state. This is related to the gap $\Delta$ of the map ${\cal E}$, cf. section \ref{RunTime}.
Hence in the end we can give an error estimate for a single application of the map, which is of the order
\begin{equation}
\epsilon^{sg} = {\mathcal O}\left(\frac{m}{2e(1-c)}\frac{4\pi}{t} \beta 2^{-r}\right).
\end{equation}

\section{Implementation} \label{implementation}
In this section we describe how to efficiently implement the quantum gates required by our algorithm on a quantum computer.
As is now standard in the literature, we assume that we can implement single-qubit operations, measurements of the
observables $\sigma^\alpha$, and elementary two-qubit gates, such as the {\sc cnot} gate with unit cost.

The first nontrivial operation required by our procedure is a means to simulate the unitary dynamics $e^{-itH}$
generated by a $k$-particle Hamiltonian $H$. We assume that $H$ can be written as the sum of $s$ terms,
each of which is easy to simulate on a quantum computer. The best way to do this follows the method described
by Berry \emph{et. al.}\ \cite{berry:2007a} and by Childs \cite{Childs:2004}: this procedure provides a simulation of the dynamics $e^{-itH}$
using a quantum circuit of length $T_H$, where

\begin{equation}
    T_H = cs^2 t_0N (\log_\ast(N))^2 9^{\sqrt{\log(s^2t_0/\epsilon_H)}},
\end{equation}

and $c$ is a constant, $s$ denotes the number of summands in $H$, $0\le t\le t_0$, $\epsilon_H$ is the desired error, and $\log_\ast(N)$ is the function defined by
$$
\log_\ast(N) \equiv \min\{r\,|\, \log_2^{(r)}(N)\},
$$
where $\log_2^{(r)}(\cdot)$ is the $r$th iterated logarithm. Now, for a typical Hamiltonian encountered in condensed matter physics or quantum chemistry, the number of terms $s$ scales as a polynomial with $N$, the number of particles. Thus the length $T_H$ of the circuit scales better than any power of $1/\epsilon_H$ and is almost linear with $t_0$ and scales slightly worse than a polynomial in $N$. Thus we can simulate $e^{-itH}$ for a length of time $t\sim p(N)$ and to precision $\epsilon_H \sim 1/q(N)$ with an effort scaling polynomially with $N$, where $p$ and $q$ are polynomials.

The next operation required by our algorithm is a method to measure the observable $H$. This can be done by making use of the quantum phase estimation \cite{kitaev:1995a,cleve:1997a}, which is a discretization of von Neumann's prescription to measure a Hermitian observable.  First adjoin an ancilla -- the \emph{pointer} -- which is a continuous quantum variable initialized in the state $|0\rangle$, so that the system+pointer is initialized in the state $|\psi\rangle|0\rangle$, where $|\psi\rangle$ is the initial state of the system. Then evolve according to the new Hamiltonian $K = H\otimes\hat{p}$ for a time $t$, so the evolution is given by
\begin{equation}
    e^{-it H\otimes \hat{p}} = \sum_{j=1}^{2^N} |\psi_{j}\rangle\langle \psi_{j}|\otimes e^{-itE_j \hat{p}}.
\end{equation}
Supposing that $|\psi\rangle$ is an eigenstate $|\psi_{j}\rangle$ of $H$ we find that the system evolves to
\begin{equation}
    e^{-it H\otimes \hat{p}}|\psi_{j}\rangle|0\rangle = |\psi_{j}\rangle |x = tE_j\rangle.
\end{equation}
A measurement of the position of the pointer with sufficiently high accuracy will provide an approximation to $E_j$.

To carry out the above operation efficiently on a quantum computer we discretize the pointer using $r$ qubits, replacing the continuous quantum variable with a $2^r$-dimensional space, where the computational basis states $|z\rangle$ of the pointer represent the basis of \emph{momentum} eigenstates of the original continuous quantum variable. The label $z$ is the binary representation of the integers $0$ through $2^r-1$. In this representation the discretization of the momentum operator becomes
\begin{equation}
    \hat{p} = \sum_{j=1}^r 2^{-j} \frac{\mathbb{I}-\sigma^z_j}{2}.
\end{equation}
With this normalization $\hat{p}|z\rangle = \frac{z}{2^r}|z\rangle$.
Now the discretized Hamiltonian $K = H\otimes \hat{p}$ is a sum of terms involving at most $k+1$ particles,
if $H$ is a $k$-particle system. Thus we can simulate the dynamics of $K$ using the method described above.

In terms of the momentum eigenbasis the initial (discretized) state of the pointer is written
\begin{equation}
    |x=0\rangle = \frac{1}{2^{r/2}}\sum_{z=0}^{2^r-1} |z\rangle.
\end{equation}
This state can be prepared efficiently on quantum computer by first initializing the qubits of the pointer
in the state $|0\rangle \cdots |0\rangle$ and applying an (inverse) quantum Fourier transform.
The discretized evolution of the system+pointer now can be written
\begin{equation}
    e^{-it H\otimes \hat{p}}|\psi_{j}\rangle|x=0\rangle = \frac{1}{2^{r/2}}\sum_{z=0}^{2^r-1} e^{-iE_j zt/2^r}|\psi_{j}\rangle |z\rangle.
\end{equation}
Performing an inverse quantum Fourier transform on the pointer leaves the system in the state $|\psi_{j}\rangle\otimes|\phi\rangle$, where
\begin{equation}
    |\phi\rangle = \sum_{x=0}^{2^r-1} \left( \frac{1}{2^{r}}\sum_{z=0}^{2^r-1}e^{\frac{2\pi i}{2^r}\left(x-\frac{E_j t}{2\pi}\right)z} \right)|x\rangle.
\end{equation}
Thus we find that
\begin{equation}
    |\phi\rangle = \sum_{x=0}^{2^r-1} f(E_j, x)|x\rangle,
\end{equation}
where
\begin{equation} \label{pointerDist}
    |f(E_j, x)|^2 = \frac{1}{4^{r}}\frac{\sin^2\left(\pi \left(x-\frac{E_j t}{2\pi}\right)\right)}{\sin^2\left(\frac{\pi}{2^r} \left(x-\frac{E_j t}{2\pi}\right)\right)},
\end{equation}
which is strongly peaked near $x = \lfloor \frac{E_jt}{2\pi} \rfloor$. To ensure that there are no overflow errors we need to
choose $t < \frac{2\pi}{\|H\|}$. (We assume here, for simplicity, that $H\ge 0$.)

It is easy to see that actually performing the simulation of $K$ for $t=1$ using the method of \cite{berry:2007a}
requires a product of $r$ simulations of the evolution according to $\frac{1}{2^{r}} H\otimes \frac{\mathbb{I}-\sigma^z_k}{2}$ for $1, 2, 2^2, \ldots, 2^{r-1}$
units of time, respectively.

We write $\Phi$ for the unitary operation representing the complete quantum phase estimation procedure.
Using $\Phi$ it is straightforward to describe a procedure to approximate a measurement of $H$:
we adjoin $r$ ancilla qubits and apply $\Phi$ and then measure the ancilla qubits in the computational basis,
approximately projecting the system into an eigenstate $| \psi_{j}\rangle$ of energy, and resulting in a string
$x$ which is an $r$ bit approximation to the value $E_j/\|H\|$.

Finally, let us briefly discuss how to implement the unitary gate $W(E_k,E_i)$.
This is a single qubit  unitary conditioned on two energy registers.
That this conditional unitary can be performed efficiently follows by observing that one can
efficiently compute the angle $\theta = \arcsin(e^{\frac{\beta}{2}(\frac{2 \pi x}{t}-E_i)})$ into a scratchpad register,
conditionally rotate the answer qubit by this angle, and uncompute $\theta$.

\section{Quantum detailed balance} \label{quDetB}

In this section we discuss the implications of Quantum detailed balance.
We use detailed balance as a tool to ensure, that
the constructed quantum Markov chain has the desired fixed point.

\begin{paragraph}{Definition: Quantum detailed balance}
Let ${\cal E}$ denote a completely positive map, and let $\sigma$ be a density matrix, then the ${\cal E}$ is said to obey detailed balance
with respect to $\sigma$ if the induced map, ${\cal E}_\sigma[ \rho] \equiv {\cal E}[\sigma^{1/2} \rho \sigma^{1/2}]$
is Hermitian with respect to the Hilbert-Schmidt scalar product. That is, the map has to satisfy
$\mbox{Tr} \left [\rho^\dagger {\cal E}_\sigma[ \phi] \right] = \mbox{Tr} \left [{\cal E}_\sigma[ \rho]^\dagger \phi \right]$
for all complex square matrices  $\rho$ and $\phi$.
\end{paragraph}

If the completely positive map obeys detailed balance, we can immediately infer several properties. First of all, since ${\cal E}$ can be obtained from
${\cal E}_\sigma$ by a similarity transformation,  ${\cal E}$ must have a spectrum that is real.
Furthermore $\sigma$ is guaranteed to be a fixed point of the completely positive map:

\begin{paragraph}{Lemma: fixed point}\label{lem:steady}
Let $\sigma$ be a state and ${\cal E}[\rho] = \sum_{\mu}A_{\mu}\rho A_{\mu}^\dagger$ a completely positive map that
satisfies the definition for Quantum detailed balance with respect to $\sigma$, then $\sigma$ is the steady state of ${\cal E}$.
\end{paragraph}

\begin{paragraph}{Proof:}
Consider the two maps, ${\cal E}_{\sigma}[\rho] = \sum_\mu A_\mu \sigma^{1/2}\rho \sigma^{1/2} A_\mu^\dagger$
and ${\cal E}^*_\sigma[\rho] = \sum_{\mu} \sigma^{1/2} A_\mu^\dagger  \rho A_\mu \sigma^{1/2}$. By definition
${\cal E}_{\sigma}[\rho] = {\cal E}_{\sigma}^*[\rho]$ for all $\rho$. Then
\[{\cal E}[\sigma] = {\cal E}_{\sigma}[\mathbb{I}] =
{\cal E}_{\sigma}^*[\mathbb{I}] = \sigma^{1/2} \sum_\mu A_\mu^\dagger A_\mu \sigma^{1/2} = \sigma.\]
\end{paragraph}

We will now derive a simple criterion to verify whether a given channel is detailed balanced with respect to a specific
state. Suppose the basis in which the density matrix is diagonal is known, then the detailed balance
condition can be checked in a straightforward manner:

\begin{paragraph}{Lemma: Detailed balance criterion}
Let $\{ |\psi_i \rangle \}$ be a  complete basis of the physical Hilbert space  and let  $\{p_i\}$ be a probability distribution on this basis.
Furthermore, assume that a completely positive map ${\cal E}[\rho] = \sum_\mu A_\mu \rho A_\mu^\dagger$ obeys
\begin{equation}\label{criterion}
\sqrt{p_np_m}  \langle{\psi_i} |{\cal E}[|{\psi_n}\rangle \langle {\psi_m}|]|{\psi_j} \rangle
 = \sqrt{p_ip_j}   \langle{\psi_m} |{\cal E}[|{\psi_j}\rangle \langle {\psi_i}|]|{\psi_n} \rangle,
\end{equation}
then $\sigma = \sum_i p_i |{\psi_i} \rangle \langle {\psi_i} |$ and ${\cal E}$ obey the detailed balance condition.
Therefore $\sigma$ is the fixed point of ${\cal E}$.
\end{paragraph}

\begin{paragraph}{Proof:}
Let ${\cal E}_\sigma$ be defined with respect to $\sigma = \sum_{i} p_i |{\psi_i} \rangle\langle {\psi_i} | $. We need to verify, whether
${\cal E}_\sigma$ becomes Hermitian with respect to the Hilbert-Schmidt scalar product. One immediately sees that
\begin{eqnarray*}\mbox{Tr}\left[\rho^\dagger{\cal E}_{\sigma}[\phi]\right] &= &\sum_{ij;nm} \overline{\rho}_{ji} \phi_{nm} \sqrt{p_n p_m}
\langle{\psi_j} |{\cal E}(|{\psi_n} \rangle \langle {\psi_m} |) |{\psi_i} \rangle \\ & =& \sum_{ij;nm}\overline{\rho}_{ji} \phi_{nm} \sqrt{p_ip_j}
\langle{\psi_m} |{\cal E}(|{\psi_i} \rangle \langle {\psi_j} |) |{\psi_n} \rangle\\ & =&   \mbox{Tr}[{\cal E}_\sigma(\rho)^\dagger \phi].\end{eqnarray*}
\end{paragraph}

\section{Binary measurements and pairs of subspaces}\label{subSpace}
The key technical reason why it is possible to implement the reject move in the quantum Metropolis
algorithm is related to a very special normal form in which two (non-commuting) Hermitian projectors can be brought.

\begin{paragraph}{Lemma: Jordan 1875}
Let $P_1$ and $Q_1$ be two projectors of $\text{rank}(Q_1) = q$ and $\text{rank}(P_1) = p$ on a Hilbert space ${\cal H} = \mathbb{C}^n$ with $p + q \leq n$.
We assume w.l.o.g., that $q \geq p$. Then there exists a basis of ${\cal H}$ in which $P_1$ and $Q_1$ can be written in the form
\begin{eqnarray}\label{LemJordan}
&&P_1  = \left( \begin{array}{cc} \mathbb{I}_p & 0_{n-p,p} \\  0_{p,n-p} & 0_{n-p,n-p} \end{array} \right) \\\nonumber \vspace{0.1cm}
&&Q_1 = \left( \begin{array}{cccc} D_p & \sqrt{D_p(\mathbb{I}_p -D_p)} & 0 & 0 \\ \sqrt{D_p(\mathbb{I}_p -D_p)} & \mathbb{I}_p -D_p & 0 & 0 \\
0 &  0 &\mathbb{I}_{q-p} & 0 \\ 0 &  0 & 0 &0_{n-(q+p),n-(q+p)}  \end{array} \right).
\end{eqnarray}
Here, $D$ is a $p \times p$ diagonal matrix with real entries $0 \leq d_1 \leq \ldots \leq d_p \leq 1$.
\end{paragraph}

\begin{paragraph}{Proof:}
We can always choose a basis of ${\cal H}$ in which the projector $P_1$ can  be written as
\begin{eqnarray}
P_1  = \left( \begin{array}{cc} \mathbb{I}_p & 0_{n-p,p} \\  0_{p,n-p} & 0_{n-p,n-p} \end{array} \right).
\end{eqnarray}
In any basis, a  general rank $q$ projector $Q_1$ can be written in the form
\begin{eqnarray}
Q_1 = \left(\begin{array}{c} A_{pq}  \\  B_{n-p,q}\end{array} \right)  \left( \begin{array}{c c} A_{pq}^\dagger & B_{n-p,q}^\dagger \end{array} \right)
\end{eqnarray}
Here $A_{pq}$ and $B_{n-p,q}$ are rectangular matrices over $\mathbb{C}$.  We require that $Q_1$ is a projector:
$Q_1^2 = Q_1$ leads to the constraint
\begin{eqnarray} \label{projCond}
 A_{pq}^\dagger A_{pq}  +  B_{n-p,q}^\dagger B_{n-p,q} = \mathbb{I}_q.
\end{eqnarray}
We can now choose to perform a singular value decomposition of $A_{pq} = U_A \Sigma_A V_A^\dagger$ and $B_{n-p,q} = U_B \Sigma_B V_B^\dagger$.
The projector can thus be written as
\begin{eqnarray}
Q_1 =  \left( \begin{array}{c c} U_A & 0 \\ 0 & U_B \end{array} \right)
\left(\begin{array}{c c }   \Sigma_A\Sigma_A^\dagger  & \Sigma_A V_A^\dagger V_B \Sigma_B \\  \Sigma_B V_B^\dagger V_A \Sigma_A & \Sigma_B \Sigma_B^\dagger \end{array} \right)
\left( \begin{array}{c c} U_A^{\dagger} & 0 \\ 0 & U_B^\dagger \end{array} \right).
\end{eqnarray}
Note that $U_A$ and $U_B$ are $p$- and $(n-p)$-dimensional unitary matrices respectively.
Therefore, the total block diagonal unitary $U_A \oplus U_B$ leaves the projector $P_1$ invariant.
If we turn to equation (\ref{projCond}), we see that upon inserting
the singular value decomposition, the matrix $V = V_A^\dagger V_B$ must satisfy
\begin{equation} \label{sim}
\Sigma_A^\dagger\Sigma_A = V ( \mathbb{I}_q - \Sigma_B^\dagger \Sigma_B ) V^{\dagger}.
\end{equation}
Note that both $\Sigma_A^\dagger\Sigma_A$ and $\mathbb{I}_q - \Sigma_B^\dagger \Sigma_B$ are diagonal matrices,
which are according to (\ref{sim}) similar. If we assume w.l.o.g., that the singular values are non-degenerate,
we conclude that $V$ can only be a permutation matrix. The degenerate case can be covered by a continuity argument.
If we define $D = \Sigma_A\Sigma_A^\dagger$ and apply the appropriate permutations to the remaining submatrices,
we are left with the desired expression for $Q_1$.
\end{paragraph}

To make the binary measurements complete, we have to choose the complementary projectors as $P_0 = \mathbb{I} - P_1$
and $Q_0  = \mathbb{I} - Q_1$; obviously, those complementary measurement projectors have a very similar structure to $P_1$ and $Q_1$.

\section{An experimental implementation}\label{experiImp}
\begin{figure}
\begin{center}
\scalebox{0.45}
{
\includegraphics*{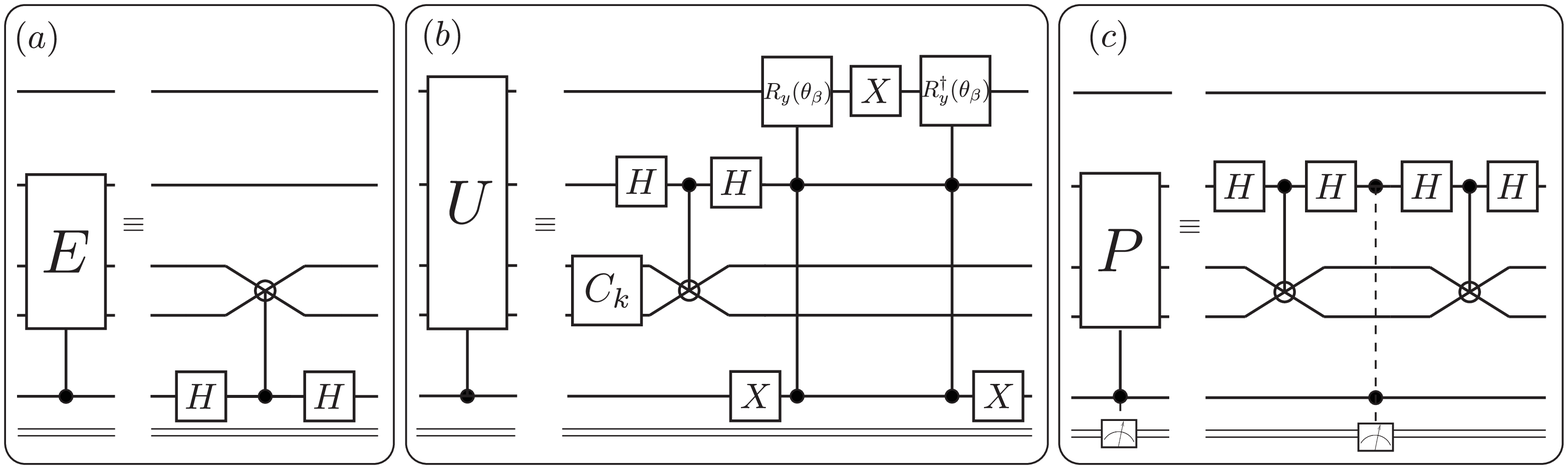}
}
\end{center}
\caption{
Fig. ~(a) describes the  first phase estimation step of the circuit. Since the phase estimation of the two-qubit Heisenberg Hamiltonian can be implemented
exactly by the Fredkin gate, a single phase estimation operation is sufficient. In Fig .~(b) the elementary unitary of the circuit is depicted. The angle
of the controlled-controlled $R_y(\theta_\beta)$ rotation needs to be chosen such that $\cos(\theta_\beta) = e^{-\beta}$. The final measurement $P$ is depicted
in Fig. ~(c). The first phase estimation has to be followed by a measurement which verifies that the two phase estimation bits are equal. The phase
estimation is then undone so that $P$ is a Hermitian projector.}
\label{Fig.exp}
\end{figure}

It is possible to implement the quantum Metropolis algorithm with todays technology for a simple 2 qubit example system.
Here, we will show how the different building blocks of the quantum Metropolis algorithm
can be represented with simple quantum circuits. For this we need to consider a quantum computer of 5 qubits.
Let us assume that we want to simulate the Gibbs states of the Heisenberg ferromagnet on 2 spin 1/2Õs, i.e.
\begin{equation}
	H_2 = -\frac{1}{2}\left( \sigma^x_1\otimes \sigma^x_2 +  \sigma^y_1\otimes \sigma^y_2 +  \sigma^z_1\otimes \sigma^z_2 \right),
\end{equation}
which is certainly one of the most interesting Hamiltonians for 2 qubits. With the appropriate energy offset,
this Hamiltonian has the spectrum $\{0, 2\}$, where the eigenvalue $0$ is threefold degenerate. This is very good
news, as it means that an exact phase estimation algorithm can be set up with just a single (qu)bit of accuracy.
Such a phase estimation requires simulating the Hamiltonian for a time $t = \pi/2$. One sees that this unitary
corresponds exactly to the $\mbox{SWAP}$ gate. That is,
\begin{equation}
U\left(\frac{\pi}{2}\right) = e^{-i\frac{\pi}{2}H_2} = \mbox{SWAP}.
\end{equation}
In the quantum Metropolis algorithm, we need to implement the controlled version of this $\mbox{SWAP}$, which
is the Fredkin gate. In \cite{Milburn}, it has been shown how this Fredkin gate can be implemented efficiently using
optics. A related gate, the so-called Toffoli gate, was recently realized in the group of R. Blatt with an ion
trap computer \cite{Blatt}. The second gate to be implemented is the controlled Metropolis unitary $W$.
The Metropolis unitary can be implemented with two controlled $R_y$ rotations:
\begin{equation}\label{expW}
	W(\theta_\beta) = R_{y}(-\theta_\beta)_C \;  X  \; R_{y}(\theta_\beta)_C,
\end{equation}
where we have made use of the standard single qubit unitary, $ R_{y}(\theta) = \exp(-i \frac{\theta_\beta}{2} \sigma^y) $ and
wrote $X = \sigma^x$. The temperature can be controlled by the angle $\theta_\beta$. Comparison with the original Metropolis unitary
(\ref{Wmatrix}) shows that we have to set $\cos(\theta_\beta) = e^{-\beta}$.  The full circuit is depicted in
 Fig.~\ref{Fig.exp}. Note that this circuit can be simplified, if we regard the lowest qubit as a classical bit, which is
 determined by the first phase estimation. It is possible to condition the remainder of the
 circuit on the first phase estimation result. Then the controlled Metropolis unitary $W$ can be implemented
 by a single $\mbox{CNOT}$ operation.\\

 Let us briefly recall the necessary steps that are needed to implement the algorithm for this
 five-qubit example crcuit. Since the phase estimation procedure is exact, the algorithm simplifies greatly and
 all assumptions for the steps described in section \ref{mainAlg} are met. We will recall the steps again in
 this paragraph, so that the section is sufficiently self-contained.  We will, however, be less general and
 focus the description on the two-qubit Heisenberg Hamiltonian.  The qubits that comprise the circuit
 are labeled according to Fig.~\ref{Fig.exp}, even though the order in which they are written
 corresponds to the notation used in the remaining part of the paper. This means that the first register, labeled by
 $| \cdot \rangle_{23}$, contains the physical state of the system from which we sample. The second register
 $| \cdot \rangle_{1}$  contains the value of the first phase estimation as indicated by the operation $E$
 (see Fig.~\ref{Fig.exp}a).  Register number three is comprised of the fourth qubit $|\cdot \rangle_{4}$ and
 is used for the second phase estimation procedure which is part of the unitary $U$   ( see Fig.~\ref{Fig.exp}b).
 Finally, the fourth and last register is given by the accept/reject qubit number 5, i.e. $|\cdot\rangle_5$.

\begin{paragraph}{Step 0:}
Initialize the full circuit to the inital state
\[
	| \psi_{0} \rangle =  | 0 \; 0 \rangle_{23}  | 0 \rangle_1  | 0 \rangle_4  | 0 \rangle_5.
\]
After the initialization continue with {\it Step 1}.
\end{paragraph}

\begin{paragraph}{Step 1:}
We currently are in a state that is of the form
\[
       | \psi_{0} \rangle = | \psi \rangle_{23}  | 0 \rangle_1  | 0 \rangle_4  | 0 \rangle_5,
\]
where $| \psi Ê\rangle_{23} $ is some arbitrary two-qubit state stored in the second and third qubit.
Apply the phase estimation map $E$ as given in  ( see Fig.~\ref{Fig.exp}a) and {\it measure qubit number $1$}.
\begin{eqnarray*}
| \psi_{0} \rangle &\rightarrow& E | \psi_{0} \rangle  = \sum_{i=1}^{4} \langle \psi_i | \psi \rangle \;\; | \psi_i \rangle_{23} | E_i \rangle_1  | 0 \rangle_4  | 0 \rangle_5
\;\;\;\;\;\;\; \mbox{apply $E$ }\\
&\rightarrow& \;\;\; | \psi_{1} \rangle =  | \psi_i \rangle_{23} | E_i \rangle_1  | 0 \rangle_4  | 0 \rangle_5.
\;\;\;\;\;\;\; \;\;\;\;\;\;\; \;\;\;\;\;\;\; \mbox{{\it measure} qubit $1$}
\end{eqnarray*}
Here the $|\psi_i\rangle$ denote the eigenvectors of the Heisenberg Hamiltonian with the energies marked by $E_i \in \{0,1\}$.
Go to {\it Step 2}.
\end{paragraph}

\begin{paragraph}{Step 2:}
We start by drawing a random unitary $C$ with respect to a uniform probability distribution from the set of Pauli matrices
$\{ \sigma^x_2 , \sigma^x_3, \sigma^z_2, \sigma^z_3 \}$ acting on either of the two qubits labeled by $2$ and $3$.
We now apply the corresponding unitary $U$ of Fig.~\ref{Fig.exp}b to the state $| \psi_1 \rangle$.
\[
	|\psi_2 \rangle = U |\psi_1 \rangle = \sum_{k=1}^4 x_{ki} \sqrt{f_{ki}} | \psi_k \rangle_{23} | E_i \rangle_1  | E_k \rangle_4  | 1 \rangle_5 +
	\sum_{k=1}^4x_{ki} \sqrt{1 - f_{ki}} | \psi_k \rangle_{23} | E_i \rangle_1  | E_k \rangle_4  | 0 \rangle_5.
\]
Here, the  $x_{ki}$ denote the matrix elements of $C$ in the eigenbasis of $H$ and $f_{ij}$ stems form the $W$ matrix (\ref{Wmatrix}), which
is implemented by (\ref{expW}) and depicted in Fig.~\ref{Fig.exp}b.\\
{\it Measure qubit number $5$}.\\\\
{\bf accept:} If  the measurement outcome is  $1$  the corresponding state is proportional to
\[
	|\psi_+\rangle \propto \sum_{k=1}^4 x_{ki} \sqrt{f_{ki}} | \psi_k \rangle_{23} | E_i \rangle_1  | E_k \rangle_4  | 1 \rangle_5.
\]
We then measure the second phase estimation register comprised of qubit number $4$. With a probability proportional to
$|x_{ki}|^2f_{ki}$ the resulting state will collapse to an eigenstate of the Hamiltonian that is of the form
\[
	 | \psi_k \rangle_{23} | E_i \rangle_1  | E_k \rangle_4  | 1 \rangle_5 .	
\]
Then go to {\it Step 4}.\\\\
{\bf reject:} Otherwise, if the measurement outcome will be $0$ and the state is proportional to
\[
 |\psi_-\rangle \propto \sum_{k=1}^4x_{ki} \sqrt{1 - f_{ki}} | \psi_k \rangle_{23} | E_i \rangle_1  | E_k \rangle_4  | 0 \rangle_5,
\]
we have to start the rejection procedure. Go to {\it Step 3}.
\end{paragraph}

\begin{paragraph}{Step 3:}
We need to reject the proposed update. To this end we have to implement the measurement scheme as indicated in Fig. ~\ref{Fig.walk}.
The first thing we need to do is to apply the adjoined $U^\dagger$ of the unitary  in Fig.~\ref{Fig.exp}b to $|\psi_-\rangle$. We are
left with the state
\[
	| \psi_{rej} \rangle = U^\dagger |\psi_-\rangle.
\]
Starting from this state, we implement the following measurement scheme.
\begin{center}
{\it Measure the projector $P_s$ }  (Fig.~\ref{Fig.exp}c),
\end{center}
where the outcome $s = 1$ corresponds to the case where the two energies agree and the outcome $s = 0$ to the
the case where two energies disagree. \\

{\bf success $s=1$}
This outcome heralds that the energies coincide and that we successfully returned to the state prior to the proposed update $C$.
Hence we have returned to a state in the energy $E_i$ subspace. Go to {\it Step 4}.\\

{\bf failure $s=0$}
We have failed to return to the original energy subspace. To unwind the state and to return to the original state we have to introduce
a further binary projective measurement $Q_s$. The measurement is related to the unitary $U$ in the following manner. First
we apply $U$ again, then we {\it measure qubit  $|\cdot\rangle_5$}, and finally we apply $U^\dagger$. Hence the $Q_s$ measurement reads
\[
	Q_s = U^\dagger \;  \I_{23} \otimes \I_{1} \otimes \I_{4} \otimes |s\rangle\langle s |_5 \; U.
\]
We now have to alternate the measurements $Q_s$ and $P_s$. That is, we now repeatedly apply $Q_s$, disregard the measurement
outcome and apply $P_s$,
\[
	Q_s \rightarrow \left. P_s \right |_{s=0} \rightarrow Q_s \rightarrow \left. P_s \right |_{s=0} \ldots
\]
until we measure the projector $P_1$ once. The corresponding plan of action is given by Fig. ~\ref{Fig.walk}.
The result $P_1$ indicates, that we have successfully returned to the original energy subspace.  Go to {\it Step 4}.
\end{paragraph}

\begin{paragraph}{Step 4:}
To finalize the single application of the Metropolis rule, we have to clean up the ancilla registers and prepare them for a subsequent
application. The current state of the system is of the form
\[
 	| \psi_k \rangle_{23} | E_i \rangle_1  | E_k \rangle_4  | s \rangle_5.	
\]
This state has to be mapped to
\[
 	| \psi_k \rangle_{23} | E_i \rangle_1  | E_k \rangle_4  | s \rangle_5	\rightarrow | \psi_k \rangle_{23} | 0 \rangle_1  | 0 \rangle_4  | 0 \rangle_5.
\]
This prepares a new input state for the subsequent application of the Metropolis rule. We now have to return to {\it Step 1}.
\end{paragraph}

 This completes the description of the Metropolis algorithm. The number of times the sequence of the Steps 1 -- 4 has to be
 repeated  before a valid sample is prepared is related to the mixing time of the algorithm ( see section \ref{RunTime}).
 Note, that a different choice for the unitaries $\{ C \}$ in Step 2 is also possible.  The only requirements are that
 the probability of applying $C$ is equal to that of applying $C^\dagger$, and that the updates allow transitions between
 all the eigenstates.

\section{Simulation of quantum many-body systems}
It would go far beyond the scope of this paper to give a faithful account on only the most
eminent applications of the quantum Metropolis algorithm to the simulation of quantum many-body systems.
We will therefore give only a brief sketch in this section on how we expect that the devised quantum algorithm
will aid in the computation of static properties of some notoriously hard problems in quantum physics
that have eluded direct computation for large system sizes by classical means. Such problems are for
instance the determination of the phase diagram of the Hubbard model, the computation of binding energies of complex molecules
in quantum chemistry, and the determination of the hadron masses in gauge theories. Common to these problems is that
the particles are strongly interacting fermions and bosons. We expect that it is this class of problems where our algorithm will
be able to give the strongest contributions. At this point we would like to point out, that the quantum Metropolis algorithm is
not plagued by the notorious sign problem, because the algorithm allows one to sample directly in the eigenbasis of the Hamiltonian.
This can be done irrespectively of whether the degrees of freedom are bosonic or fermionic.\\
In order to implement the quantum Metropolis algorithm for a specific many-body Hamiltonian $H$, we need to be able
to perform the phase estimation algorithm efficiently. The central subroutine that needs to be implemented is therefore
the simulation of the time evolution for the Hamiltonian  $H \otimes \hat{p}$, as was discussed previously in section
\ref{implementation}. The simulation method described in \cite{berry:2007a}
relies on the fact that we are able to decompose the Hamiltonian into a sum of local Hamiltonians $h_k$ with
$H = \sum_k h_k$ that can be simulated by themselves on a quantum computer efficiently.
A method to rephrase fermionic or bosonic degrees of freedom in terms of the quantum computational degrees of freedom ,
that is in terms of qubits, is therefore needed. Such a program was devised in \cite{abrams:1997,Ortiz1,Ortiz2}
and we merely give a brief overview here and refer the reader to the corresponding references.

\begin{figure}
\begin{center}
\scalebox{0.5}
{
\includegraphics*{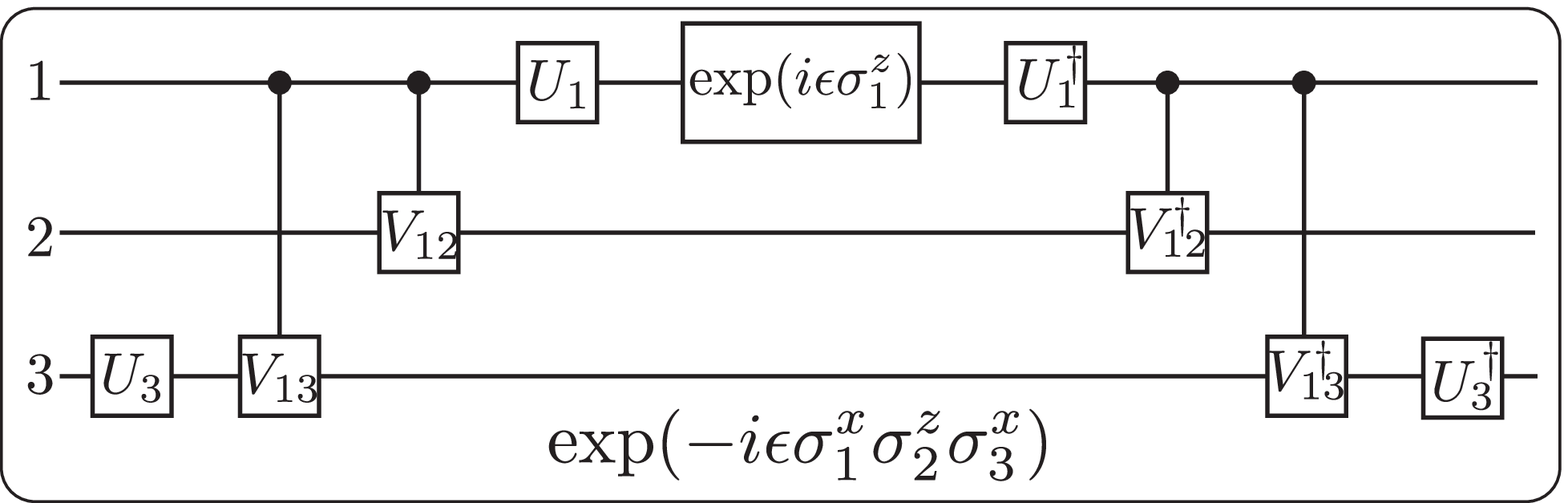}
}
\caption{A fermionic many particle Hamiltonian can be simulated on a quantum computer by mapping the fermionic degrees of freedom to spin-1/2 particles
\cite{Ortiz1,Ortiz2}. Such a mapping is given by the famous Jordan-Wigner transformation. Here, the fermionic algebra can be expressed in terms of the $su(2)$ algebra via
$c_k^{\dagger} = -\left(\otimes_{l=1}^{k-1} \sigma^z_l \right) \sigma^{+}_k$, where $\sigma^+_k = \frac{1}{2}\left(\sigma^x_k + i \sigma^y_k\right)$.
The dynamical part of the fermionic many-body Hamiltonian often contains terms of the form $h_{kj} = c_k^\dagger c_j + c_j^\dagger c_k$,
which become non-local after the transformation. Operators that are not adjacent in terms of the labeling often contain a chain of Pauli $\sigma^z$ operators in between them.  A typical term of this kind that occurs after this transformation is $h^{X}_{kj} = \sigma^x_k(\otimes_{l=k+1}^{j-1}\sigma^z_l)\sigma^x_j$. To simulate the time evolution of such a non-local term on a quantum computer, we need to be able to decompose this unitary into two qubit gates. Given the two unitaries $V_{kl} = \exp( i \frac{\pi}{4}\sigma^z_k\sigma^z_l)$ and $U_l = \exp(i \frac{\pi}{4}\sigma^y_l)$ such a decomposition is indeed possible as depicted in the
above circuit for the evolution of $\exp(-i\epsilon \sigma^x_1 \sigma^z_2 \sigma^x_3)$.}
\label{figFermTrot}
\end{center}
\end{figure}

\paragraph{The Hubbard model:}
The Hubbard model \cite{Hubbard} is based on a tight binding approximation that
describes electrons in a periodic potential confined to move only in the lowest Bloch band.
The Hubbard Hamiltonian consists of a hopping term and an interaction term written in form of fermionic creation $c_{i,\sigma}^\dagger$
and annihilation $c_{i,\sigma}$ operators that act on a lattice site $i$ in a regular lattice of $N$ sites.
\begin{equation}
H = -t \sum_{<i,j>,\sigma} \left(c_{i,\sigma}^\dagger c_{j,\sigma} + c_{j,\sigma}^\dagger c_{i,\sigma}\right) + U \sum_i n_{i,\downarrow}n_{i,\uparrow}
\end{equation}
This Hamiltonian has to be expressed in terms of spin degrees of freedom in order to be implemented in the standard
quantum circuit formulation. The interaction term can be seen to be implementable directly since the particle density $n_{i,\sigma}$
operator acts only locally and is bosonic in nature. The implementation of the hopping term is a bit more challenging. Consider for simplicity
the hopping term for a single electron spin only. This part can be expressed in terms of the Jordan-Wigner transformation, cf. Fig. \ref{figFermTrot}, as
\begin{equation}
	t \sum_{<i,j>} \frac{1}{2}\left(\sigma^x_i (\otimes_{k = i+1}^{j-1} \sigma_k^z)\sigma^x_j
	+ \sigma^y_i (\otimes_{k = i+1}^{j-1}\sigma_k^z)\sigma^y_j \right),
\end{equation}
once a specific order of the $N$ lattice sites has been chosen. As is shown in Fig. \ref{figFermTrot} the unitary evolution of each individual
summand can be implemented with a cost that scales at most linearly with the total system size \cite{Ortiz1,Ortiz2}.
More general fermionic Hamiltonians can be implemented in a similar fashion.

\paragraph{Quantum chemistry}

A central problem in Quantum chemistry is the determination of molecule properties. The major challenge is the determination of the electron binding
energies that need to be computed in dependence of the nuclei position. The general approach to this problem is to solve the approximate
Hamiltonian of the electronic degrees of freedom that arises due to the Born-Oppenheimer approximation. In this approximation the nuclei positions
are external parameters in the electronic Hamiltonian. The calculation of the molecule properties relies on the fact that the electronic energy can be
determined efficiently in dependence of the nuclei position. In their paper \cite{Kassal}, Kassal and  Aspuru-Guzik show how a quantum computer could
be used to determine molecule properties at a time that is a constant multiple of the time needed to compute the molecular energy. The algorithm
relies on a black box that computes the molecular energy for every configuration. The quantum Metropolis method can function as
this black box algorithm, which was missing so far. For the Metropolis algorithm to work, one needs to implement the phase estimation
procedure for the chemical Hamiltonian of interest.  It is shown in \cite{Guzik}, that the phase estimation procedure can be implemented efficiently
for a general second quantized chemical Hamiltonian.

\paragraph{Gauge theories}
The current most common non-perturbative approach to QCD is Wilson's lattice gauge theory \cite{Wilson}, which maps the problem to one of
statistical mechanics, where the Euclidean action now assumes the role of a classical Hamilton function.
It is therefore reasonable to assume, that lattice gauge theories would also be the method of choice for
the quantum Metropolis algorithm. However, the algorithm relies on a Hamiltonian formulation of the problem. Such a formulation is given
by Kogut and Susskind's \cite{Kogut} Hamiltonian formulation of lattice gauge theories in $3+1$ dimensions.  Here the $3$-dimensional space
is discretized and put on a cubic lattice, while time is left continuous. The fermions reside on the vertices of the lattice, while the gauge degrees
of freedom are put on the links. The physical subspace is required to be annihilated by the generators of the gauge transformation, i.e. all physical
states need to satisfy Gauss's law.\\
It turns out however, that this approach seems to be very hard to implement on a quantum computer.  This is due to the fact that each of the
links carries a Hilbert space that is infinite dimensional, namely the space of all square integrable functions on the corresponding gauge
group $SU(N)$. A finite approximation to this Hilbert space therefore leads immediately to a breakdown of the underlying symmetry. \\
A different formulation of gauge theories, that does not suffer from this problem, is therefore needed. Such a formulation is given in terms
of quantum link models introduced by Horn \cite{Horn}. Brower et al. showed that QCD and in general any $SU(N)$ gauge theory
can be expressed as a quantum link model \cite{Wiese}. In the quantum link formulation the classical statistical mechanics problem is replaced
by a problem formulated in terms of quantum statistical mechanics in which the classical Euclidean action is replaced by a quantum Hamiltonian.
The central feature is that the corresponding Hilbert space of the gauge degrees of freedom at each link is now finite. It suffices that each
link of a $SU(N)$ link model carries a single finite representation of $SU(2N)$. This is achieved by formulating the problem in $4+1$ dimensions,
where the four physical dimensions correspond to the actual physical Euclidean space time, while the fifth Euclidean dimension
plays the role of an additional unphysical dimension. The $4$-dimensional Euclidean space time is discretized and lives on a cubic lattice. Furthermore, it
was shown by Brower et al. \cite{Wiese}, that the continuum limit is obtained by sending the fifth unphysical Euclidean dimension to infinity,
which corresponds to preparing the ground state of the lattice Hamiltonian. It can be seen, that the $4+1$ dimensional link
models are related to standard gauge theories in 4 dimensions via dimensional reduction \cite{dimReduction}. \\
The full Hilbert space of the $SU(3)$ gauge theory can be written as the tensor product of a $20$-dimensional Hilbert space for each link
of the lattice and the finite dimensional fermionic Hilbert space that resembles the quarks. In contrast to the standard lattice gauge theories
the configuration space of the quantum link model resembles that of quantum spin models.\\
The physical spectrum, and by that the Hadron masses, of the $4$-dimensional theory can be obtained from computing the correlation
functions in the Euclidean direction on the ground state of the $4$-dimensional lattice Hamiltonian.

\begin{paragraph}{Acknowledgements:}
We would like to thank  S. Bravyi, C. Dellago  and J. Kempe for helpful discussions and Erwin Schr\"odinger Institute for Mathematical Physics where part of this work was done. KT was supported by the FWF program CoQuS and is responsible for the main part of this work. TJO was supported, in part, by the EPSRC. KGV is supported by DFG FG 635. DP is partially funded by NSERC, MITACS, and FQRNT. FV is supported by the FWF grants FoQuS and ViCoM, by the European grant QUEVADIS and the ERC grant QUERG.
\end{paragraph}

\end{document}